\documentclass[sigconf]{acmart}
\usepackage{amsmath}
\usepackage{amsfonts}
\usepackage{graphicx}
\usepackage{multirow}
\usepackage{textcomp}
\usepackage{xcolor}
\usepackage{algorithm}
\usepackage{algorithmicx}
\usepackage[noend]{algpseudocode}

\algnewcommand{\IfInline}[1]{\State\algorithmicif\ \, #1\ \, \algorithmicthen \, }
\algnewcommand{\EndIfInline}{\unskip\ }
\def\BibTeX{{\rm B\kern-.05em{\sc i\kern-.025em b}\kern-.08em
    T\kern-.1667em\lower.7ex\hbox{E}\kern-.125emX}}

\renewcommand\footnotetextcopyrightpermission[1]{} 

\begin{document}

\title{TriPoll: Computing Surveys of Triangles in Massive-Scale Temporal Graphs with Metadata}
\thanks{\footnotesize This work was performed under the auspices of the U.S. Department of Energy by Lawrence Livermore
	National Laboratory under Contract DE-AC52-07NA27344 (LLNL-CONF-822890).   
Funding from LLNL LDRD project 21-ERD-020 was used in this work. 

This document was prepared as an account of work sponsored by an agency of the United States government. Neither the
United States government nor Lawrence Livermore National Security, LLC, nor any of their employees makes any warranty,
expressed or implied, or assumes any legal liability or responsibility for the accuracy, completeness, or usefulness of
any information, apparatus, product, or process disclosed, or represents that its use would not infringe privately owned
rights. Reference herein to any specific commercial product, process, or service by trade name, trademark, manufacturer,
or otherwise does not necessarily constitute or imply its endorsement, recommendation, or favoring by the United States
government or Lawrence Livermore National Security, LLC. The views and opinions of authors expressed herein do not
necessarily state or reflect those of the United States government or Lawrence Livermore National Security, LLC, and
shall not be used for advertising or product endorsement purposes.}

\author{Trevor Steil, Tahsin Reza, Keita Iwabuchi, Benjamin W. Priest, Geoffrey Sanders, and Roger Pearce}
\affiliation{
	\institution{Center for Applied Scientific Computing (CASC), Lawrence Livermore National Laboratory (LLNL)}
\city{Livermore} 
\state{CA}
\country{USA}}
\email{{steil1, reza2, iwabuchi1, priest2, sanders29, pearce7}@llnl.gov}

\begin{abstract}
Understanding the higher-order interactions within network data is a key objective of network science.
Surveys of metadata triangles (or patterned 3-cycles in metadata-enriched graphs) are often of interest in this
pursuit. 
In this work, we develop \texttt{TriPoll}, a prototype distributed HPC system capable of surveying triangles in massive graphs
containing metadata on their edges and vertices.
We contrast our approach with much of the prior effort on triangle analysis, which often focuses on simple triangle
counting, usually in simple graphs with no metadata.
We assess the scalability of \texttt{TriPoll} when surveying triangles involving metadata on real and synthetic
graphs
with up to hundreds of billions of edges.
We utilize communication-reducing optimizations to demonstrate a triangle counting task on a 224 billion edge web
graph in approximately half of the time of competing approaches, while additionally supporting metadata-aware
capabilities.
\end{abstract}

\maketitle

\pagestyle{plain} 

\section{Introduction}

Network scientists seek meaningful higher-order structure within their relational datasets.
Real-world relational datasets tend to have graph topology (the relational data) with structured or unstructured metadata living on vertices and/or edges.
The discovery of how topology, timing, and other non-relational data combine to form meaningful higher-order structure facilitates exploratory data analysis of network datasets,
network-based machine learning capabilities, and foundational understanding of network function.
Such pattern discovery is a computationally challenging task; massive relational datasets often require distributed storage and computing for scalable pattern analysis
strategies.

\begin{figure}[t]
\begin{center}
\includegraphics[width=3.3in]{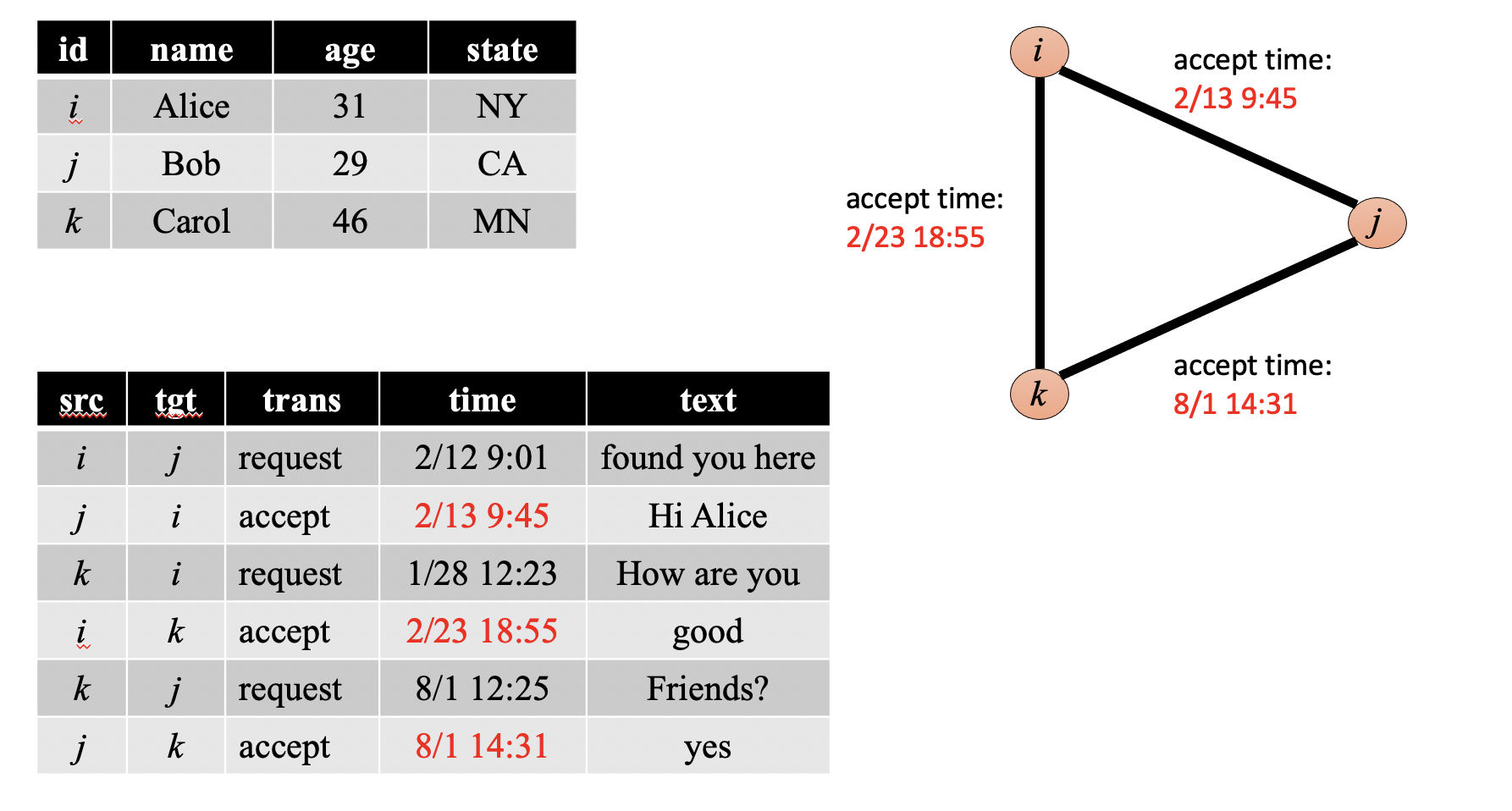}
\captionsetup{size=small}
\caption{Scenarios in a social network where analysis seeks to understand higher-order structures.
This relational dataset is organized into {\bf (Top-left)} a vertex table, listing users and their attributes, and
{\bf (Bottom-left)} an edge table, listing the times users request to connect and when connections are accepted.
{\bf (Right)}
Studying presence or absence of various higher-order structures (e.g., triangles with metadata) may aide in enhancing several network analysis tasks like link prediction
({\em Who is likely connected in real life but not in our dataset?}) and anomaly detection ({\em Which accounts are fake?}).  }
\label{fig:motivation}
\end{center}
\end{figure}

A common case we focus on here is where records contain a relationship between two vertices, a time of observation (or timestamp), and structured metadata on the vertices and edges.
This vertex and edge metadata can take the form of discrete labels and/or text fields.
Consider, for example, a large online marketplace listing interactions (edges) between users (vertices).
Vertex fields could include a user rating such as a floating point number, a discrete label such as buyer, seller, or both, and a short string username.
Meanwhile, edge fields could include a discrete type label such as message, purchase, or rating, a floating point timestamp, a floating point rating, and a possibly long message string. 
In this work we will refer to such a graph as a {\em decorated temporal graph}.  See the example in Fig.~\ref{fig:motivation}.

Triangles (graph 3-cycles) are fundamental patterns
that are commonly difficult to enumerate in massive distributed real-world networks \cite{kepner2017summary, kepner2019summary}.
These difficulties arise because networks associated with data analysis have scale-free topology
(heavy-tail degree distribution, small effective diameter, and the expander graph property \cite{hoory2006expander})
complicating many aspects of serial and distributed computation.
Often, techniques that approximate triangle counts suffice for an application, and processing every triangle is not necessary \cite{Hasan2018review}.
However, when an analysis is performed involving {\em metadata triangles} (3-cycles involving specific categories
involving labels, topics, or timings) and there are a wide variety of metadata triangles,
processing every single triangle may be required.
Additionally, processing every triangle is of interest when metadata triangle incidence upon vertices, edges, or subgraphs inform feature vectors in downstream machine learning models.

In this work we design \texttt{TriPoll}, a distributed, asynchronous graph framework that affords the exploration of the types of triangles that exist within massive datasets.
Our prototype system allows users to answer hypotheses regarding the relevance of metadata triangles on their datasets and leverage their discoveries for
automated analyses (see scenarios in Fig~\ref{fig:motivation}).

Most previous works use \emph{triangle counting} as the final goal, as  the number of triangles can be much larger
than the edge set of a massive graph.  This is often global counting (returning a single integer for the number of triangles in the entire graph) 
or local counting of triangle participation at vertices and/or edges.   Several applications make use of the latter, including performing truss decomposition \cite{cohen2008truss}, 
enhancing community detection \cite{Berry2011ToleratingTC}, 
bounding the sizes of cliques \cite{wolf2015},
computing clustering coefficients \cite{Arifuzzaman.2019.TKD.10.1145/3365676}, and 
using the local triangle counts in downstream machine learning including vertex role analysis \cite{Henderson2012RolX}.
In some cases, the goal
is to find triangles of a particular type within labeled or temporal graphs.
Work exists leveraging local vertex/edge participation in various metadata triangle types for more customizable community detection in labeled graphs \cite{Benson163}, 
as well as dynamic graphs \cite{benson2017temporal}.
Moreover, 
\cite{reza2020indexing} uses local edge participation in various metadata triangle types for performance gains in interactive labeled pattern matching. 
As many relational datasets are highly decorated with many fields at every vertex and edge, 
these works are just scratching the surface of what network scientists could ask of their datasets regarding 
metadata triangles and how the metadata triangles can be leveraged for various discovery tasks.  

This provides us with one unique capability of \texttt{TriPoll}: our users provide a {\em callback function} to perform on 
the metadata associated to the respective vertices and edges as each triangle is found. This allows the same code to be used for
counting of simple or metadata triangles in a decorated temporal graph, full triangle enumeration if desired, or more interesting data
collection that is more in line with the queries a network scientist wants answered during data exploration.

It is worth noting \texttt{TriPoll} identifies all triangles in a graph and applies the user's callback to the metadata of
each one as they are discovered. The end goal is not to search for particular patterns of vertex and edge metadata
within a graph, but instead to allow for custom surveys of triangles in a graph. A pattern matching approach would
answer the question "Where can I find all triangles involving 1 red vertex and 2 blue vertices?", whereas we seek
answers to questions of the form "Of the triangles with 1 red vertex and 1 blue vertex, what is the distribution of
colors of the third vertex?"

Additionally, \texttt{TriPoll} has no output in the traditional sense. We rely on the user-defined callbacks to have the effects
desired for output, whether that is incrementing counters to be read later, writing information on individual triangles
out to file, or preparing data for use in a downwind application.

Our work builds on \texttt{YGM} \cite{ygm}, our recently released open source asynchronous communication library, to allow for scalable
performance. \texttt{YGM} leverages buffering techniques and message serialization to allow truly
asynchronous computations where messages of heterogeneous types can be exchanged simultaneously. Using \texttt{YGM}, we are able to build vertex-centric graph data structures that are able to handle
metadata and can easily express the communication necessary for identifying triangles in a graph.
\texttt{TriPoll} provides an example application for performance and usability considerations as we further
develop \texttt{YGM}.

The research contributions made in this work are:
\begin{itemize}
	\item We introduce \texttt{TriPoll}, a scalable \texttt{C++} framework for triangle processing on graphs with metadata that, through
		user-defined callback functions for executing on the edge and vertex metadata of each triangle identified,
		allows the computation of customizable metadata triangle surveys on massive graphs.
	\item We showcase new capabilities in \texttt{YGM}, our asynchronous communication library\footnote{available at
		\url{https://github.com/LLNL/ygm}}, that builds on message
		buffering techniques seen in previous libraries \cite{Priest2019Grapl, maley2019conveyors} and adds a
	serialization layer to allow messages of heterogeneous and complicated types, e.g., \texttt{C++} strings and STL
	containers, to be exchanged simultaneously.
	\item We implement an optimization method allowing choice of direction for sending adjacency information during
		triangle identification to reduce communication.
	\item We evaluate performance on real-world and synthetic graphs with up to 224 billion edges. As a subset of
		the functionality offered by \texttt{TriPoll}, we are able to perform simple triangle counting to compare against
		previous work tailored to triangle counting. We find \texttt{TriPoll} gives comparable or better performance than
		these works, including counting triangles in a 224 billion edge web graph on 256 compute nodes in just over half of the time required
		by the only openly available code able to solve this problem.
	\item We use \texttt{TriPoll} to survey the distribution of triangle closing times seen in a temporal graph derived from
		Reddit including 9.4 billion edges with timestamps.
\end{itemize}

\section{Relation to External Work}
\label{section:related_work}

Triangle counting is a widely studied graph algorithm. With the rapid growth of the scale of real-world datasets, the demand for scalable, high-performance solutions has been growing as well. 
From the earlier parallel node-iterator~\cite{Schank2007_1000007159} family of algorithms, in recent years, a number of solutions have been developed targeting multi-core CPUs~\cite{Zhang.2018.HPEC.SharedTriangle.8547569}, 
GPUs~\cite{Wang.2019.HPEC.GPUTriangle.8916434}, (heterogeneous) distributed platforms~\cite{pearce2017triangle}, as well as FPGAs~\cite{Huang.2018.HPEC.FPGATriangle.8547536}. 
Here, we focus on related work that targets distributed platforms. 

{\bf Vertex-Centric Frameworks.} Popular vertex-centric graph frameworks, often optimized for low-complexity traversal
algorithms, have been shown to support triangle counting: GraphLab~\cite{Gonzalez:2012:PDG:2387880.2387883}, 
Galois-Gluon~\cite{Dathathri.2018.PLDI.Gluon.10.1145/3192366.3192404}, HavoqGT~\cite{Pearce:2014:FPT:2683593.2683654}, GraphMat~\cite{Sundaram:2015:GHP:2809974.2809983}, Giraph~\cite{Giraph.001}, 
GraphX~\cite{Gonzalez:2014:GGP:2685048.2685096}, Oracle PGX.D~\cite{Hong:2015:PFD:2807591.2807620} and Chaos~\cite{Roy:2015:CSG:2815400.2815408} (the core processing engine is edge-centric) are 
some of the most well-known systems. Since these frameworks are designed to support a wide variety of graph problems, while they can scale to large datasets, typically they are unable to offer algorithm-specific optimizations.

{\bf General Pattern Matching Tools.} 
Since triangle counting is also a primitive subgraph matching problem, distributed subgraph graph mining solutions such as Arabesque~\cite{Teixeira:2015:ASD:2815400.2815410}, QFrag~\cite{Serafini:2017:QDG:3127479.3131625}, 
Fractal~\cite{Dias.2019.SIGMOD.10.1145/3299869.3319875}, PruneJuice~\cite{Reza:2018:PPT:3291656.3291684}, TriAD~\cite{Gurajada:2014:TDS:2588555.2610511}, G-Miner~\cite{Chen:2018:GET:3190508.3190545} and 
GraphFrames~\cite{Dave:2016:GIA:2960414.2960416}, and  graph databases such as Neo4j~\cite{neo4j.property.graph}, OrientDB~\cite{Orientdb.001} and TigerGraph~\cite{TigerGraph.001} seamlessly support triangle matching queries; 
however, they are aimed at more general pattern matching and are unable to match the scale and performance offered by tailored triangle counting systems. 

{\bf Tailored Triangle Counting.} 
Depending on the target architecture (i.e., CPU, GPU or heterogeneous), distributed triangle counting solutions embrace different graph partitioning~\cite{Tom.2019.ICPP.DistributedTriangle.10.1145/3337821.3337853} and 
load balancing~\cite{Pandey.2021.TRUST.9373989} techniques (distributed~\cite{Ghosh.2020.HPEC.Tric.9286167} vs fully/partially replicated~\cite{Pandey.2021.TRUST.9373989}), preprocessing and/or pruning strategy (e.g., degree-based 
sorting, wedge identification~\cite{pearce2017triangle}), computation model (i.e., vertex/edge-centric~\cite{Hoang.2019.HPEC.DistTC.8916438} or linear algebra~\cite{Acer.2019.HPEC.DistributedTriangle.8916302}).
In the most expensive operation in a triangle counting kernel, i.e., computing the intersection of adjacency lists,
three fundamental approaches are commonly used, namely, binary search, merge-path, and hashing (of a vertex's 
neighborhood)~\cite{Pandey.2021.TRUST.9373989}. 

Early examples of distributed triangle counting were MapReduce implementations of node-iterator algorithms
\cite{cohen2009graph, Suri:2011:CTC:1963405.1963491}. These use a degree-ordered, 
2D graph partitioning technique. The works \cite{pearce2017triangle, Pearce.2019.HPEC.DistributedTriangle.8916243} embrace asynchronous MPI communication; a preprocessing step 
iteratively prunes degree-one vertices, orders vertices by degree, 
followed by querying wedges for closure to detect presence of triangles.   This work showed scalability using a
224 billion edge real-world webgraph on $\sim$9K cores, 
one of the largest scale to date. The work \cite{Tom.2019.ICPP.DistributedTriangle.10.1145/3337821.3337853} utilizes a 2D cyclic decomposition of the data graph; its processing resembles Cannon's parallel matrix-matrix 
multiplication algorithm and uses hashing for adjacency list intersection. Both
\cite{Arifuzzaman.2019.TKD.10.1145/3365676} and \cite{Kanewala.2018.PASC.10.1145/3218176.3218229} use 1D graph 
partitioning, and \cite{Arifuzzaman.2019.TKD.10.1145/3365676} demonstrated that partially replicating partitions offers communication efficiency and subsequently showed application of their technique to clustering coefficient computation. 
\cite{Kanewala.2018.PASC.10.1145/3218176.3218229} combined shared memory and distributed memory optimization techniques
to offer improved distributed memory performance; similar to~\cite{pearce2017triangle}, 
this solution operates on precomputed wedges; and employs a message aggregation technique to avoid latency-prone short messages. TriC~\cite{Ghosh.2020.HPEC.Tric.9286167} is a distributed memory solution that exploits the knowledge about 
the graph structure to improve inter-node communication; it creates edge-balanced partitions, performs parallel edge enumeration to identify triangles; accompanied by a batch-oriented scalable communication substrate. The authors demonstrated 
scalability using up to 8K cores on the Cori Supercomputer on up to 3.6B edge real-world graphs. \cite{Acer.2019.HPEC.DistributedTriangle.8916302} presents a linear algebra-based solution which utilizes 2D partitioning at the 
cluster level, and 1D partitioning at the node level, and achieves high throughput using MPI and Cilk for parallelism. Scalability was demonstrated using up to 1K Cores on a 3.6B edge graph. Another linear algebra-based solution is 
presented in~\cite{Yasar.2019.HPEC.Triangle.8916233}, which offers heterogeneous (CPU-GPU) processing on multi-GPU platforms like the Nvidia DGX.

TriCore~\cite{Hu.2018.SC.TriCore.8665796} targets GPU clusters and exploits streaming partitions to accommodate large
graphs. TriCore follows an edge-centric computation model and uses binary search for adjacency list intersection (which can improve 
memory coalescing on GPUs). DistTC's~\cite{Hoang.2019.HPEC.DistTC.8916438} design is comparable to that of TriCore, however, it uses static graph partitions consisting of vertex replicas, which drastically reduces the volume of inter-node 
communication. Using up to 32 GPUs, authors showed that DistTC comfortably outperforms TriCore.
H-Index~\cite{Pandey.2019.HPEC.H-INDEX.8916492} offers innovations for hashing-based adjacency list intersection on GPUs; the technique 
enables search space pruning leading to improved throughput. TRUST~\cite{Pandey.2021.TRUST.9373989}, which targets
distributed GPUs, also uses a hashing-based solution for computing  adjacency list intersection. Unlike TriCore and DistTC, 
TRUST embraces vertex-centric processing on GPUs. It follows a hashing-based 2D partitioning (also performs full/partial graph replication) and offers load balancing through a collaborative workload partitioning technique. 
TRUST \cite{Pandey.2021.TRUST.9373989} has 
demonstrated the largest scale among distributed GPU-based solutions: up to 1,000 GPUs on the Summit Supercomputer (using up to 42B edge real-world graphs).

\section{Preliminaries}
\label{sec:preliminaries}

Throughout this work, we will use $\mathcal{G}(\mathcal{V}, \mathcal{E})$ to denote a graph with vertex set $\mathcal{V}$ and edge
set $\mathcal{E} \subset \mathcal{V} \times \mathcal{V}$. Associated with any vertex $v \in V$ is metadata $meta(v)$ (containing vertex labels, strings, and/or floating point values),
and similarly we have edge metadata $meta(u,v) = meta(v, u)$ for any $(u, v) \in \mathcal{E}$ (containing edge labels, timestamps, strings, and/or floating point values). 
We assume input graphs $\mathcal{G}$ are
\emph{undirected}, that is $(u, v) \in \mathcal{E} \Leftrightarrow (v, u) \in \mathcal{E}$. For a vertex $v \in
\mathcal{V}$, we define its \emph{adjacency list} $Adj(v) = \{(v, u) \in \mathcal{E}\}$. We then let $d(v) = |Adj(v)|$ be its \emph{degree}.

A \emph{triangle} (or 3-cycle) is a 3-way interaction between vertices $p,q,r\in\mathcal{V}$,
defined by the existence of edges $(p, q), (p, r), (q, r) \in \mathcal{E}$ (and all reciprocal
edges). This triangle will be represented by $\Delta_{pqr}$. We let $\mathcal{T}( \mathcal{G} )$ be the set of
all triangles present in $\mathcal{G}$.

Vertex degree is a natural ordering for computational savings in triangle counting, as it tends to turn most or all of a high
degree vertex's undirected edges into directed \emph{in-edges} \cite{cohen2009graph}. We will use $<_+$ as
our comparison operator between vertices according to degree. That is, $u <_+ v$ if and only if
\begin{itemize}
	\item $d(u) < d(v)$, or
	\item $d(u) = d(v)$ and $hash(u) < hash(v)$ 
\end{itemize}
where we have chosen a deterministic function $hash$ to use in breaking ties.

We define an augmented graph known as the \emph{degree-ordered directed graph} \cite{cohen2009graph} \emph{(DODGr)}, denoted as $\mathcal{G}_+$,
to be the directed graph obtained by changing the pair of edges $(u, v), (v, u) \in \mathcal{E}$ into the
single directed edge $(u, v)$ such that $u <_+ v$. 

With a fixed ordering $p <_+ q <_+ r$ on $\mathcal{G}_+$, $\Delta_{pqr}$ is uniquely identifiable, as none of the alternative vertex orderings occur. 
See the left part of Fig.~\ref{fig:push}.
Throughout this work, we will canonically use $p, q$, and $r$ as the vertices on
individual triangles with $p <_+ q <_+ r$, and $p$ is referred to as the {\em pivot} (or anchor) vertex.

For a vertex $v \in
\mathcal{V}$, we use $Adj_+(v)$ as the adjacency list containing $v$'s out-edges in $\mathcal{G}_+$, and we use
$d_+(v) = |Adj_+(v)|$ as the out-degree of $v$.

A triangle $\Delta_{pqr}$ exists if and only if
we are able to identify the directed edges $(p,q), (p, r)$, and $(q, r)$ in the $\mathcal{G}_+$. 
As a first step, we can identify the pair of edges $(p, q)$ and $(p, r)$, and we must then check for the closing edge $(q, r)$. This \emph{wedge
check} is the basic unit of work in triangle enumeration, and we let $|\mathcal{W}_+|$ be the number of such wedges in $\mathcal{G}_+$.

For a triangle $\Delta_{pqr}$, we will use the shorthand $meta(\Delta_{pqr})$ to mean the six pieces of metadata
associated to $\Delta_{pqr}$'s vertices and edges.

Lastly, we use $Rank(u)$ to denote the MPI rank responsible for storing the associated adjacency lists $Adj(u)$ or
$Adj_+(u)$ (depending on context) and metadata 
$meta(u)$ and $meta(u, v)$ for all $v \in Adj_+(u)$, and all computations corresponding to $u$.

\section{Distributed Triangle Processing}

In this section, we will describe our algorithm for identifying triangles. During triangle identification,
\texttt{TriPoll} must ensure the correct vertex and edge metadata is gathered on the correct process for use in callbacks
on discovered triangles; this data cannot be simply read after a triangle is found as it is likely to be scattered
across multiple processes.

\texttt{TriPoll} uses our recently-developed open source, asynchronous communication mechanism built on MPI \cite{ygm} to handle sending messages between ranks.
We use custom distributed data structures as building blocks for our graph storage. Our communication
library is designed to handle irregular communication patterns. Triangle processing in graphs provides a
communication-intensive problem for us to assess performance and usability when integrated into a full
application.

In this section we will describe our communication schema, discuss our graph storage, and explain our distributed
vertex-centric, merge-path based algorithm in detail, and give examples of analyses that can be performed through
defining triangle callbacks.

As mentioned in Sec.~\ref{sec:preliminaries}, we consider all graphs to be undirected. Since our
algorithms operate on $\mathcal{G}_+$, a directed version of the original graph, instead of $\mathcal{G}$ itself,
this approach could be used for directed graphs as well. In the directed case, our augmented graph would be the
original graph with many edges having their directionality reversed and any bidirectional edges having one direction
removed. Additionally, each directed edge in the augmented graph may need an additional two bits of storage to give
the original directionality (as-seen, reversed, or bidirectional) for use in the user callback if the directionality
is important for an application.

\subsection{Asynchronous Communication using \texttt{YGM}}
\label{sec:comm}

In designing software to survey metadata triangles in massive graphs, two major challenges arise. 
The first is the presence of highly non-uniform communication patterns, which confounds the scalability of traditional HPC codes and also 
occurs in simple triangle counting workflows.
The second is the handling of vertex and edge metadata while affording the user the flexibility to execute arbitrary callback functions when 
surveying triangles, which requires careful data management to maintain efficiency. 

We meet these challenges with \texttt{YGM}, our asynchronous communication library built on top of MPI \cite{ygm}.
To achieve scalability, we leverage an asynchronous
communication scheme and message buffering. To grant the expressiveness necessary for our 
triangle surveys, we use remote procedure call (RPC) semantics, a collection of pre-built storage containers to
perform many of the basic tasks we frequently face, and serialize messages and function 
arguments before sending them through MPI to allow any serializable type to be easily sent to remote MPI ranks.

\subsubsection{Message Buffering}
Like many graph algorithms, na\"ive triangle enumeration in distributed memory generates a large number of small messages
as compute nodes transmit small amounts of adjacency information.
As each MPI data message creates additional overhead in the form of bits of header information and handshake protocol 
messages, such graph analysis workflows result in poor bandwidth utilization and slow wall time compared to more traditional MPI workflows.

Rather than relying upon application-specific message engineering in applications, \texttt{YGM} opaquely buffers small messages
until the buffer reaches a size threshold or is directed to flush.
By aggregating all of the small buffered messages destined for a next destination into a single large message, we can 
dramatically reduce the overhead of a na\"ive workflow without adding design complexity to the application.
Although some individual messages may take slightly longer to reach their destinations, buffering strategies of this sort
have been shown to improve overall latency in benchmark experiments \cite{Priest2019Grapl, maley2019conveyors}.

\subsubsection{Serialization}
MPI was designed for numerically-typed data in traditional HPC applications rooted in physical simulations. 
As the community continues to use MPI to solve problems in the data science realm, the limitations imposed by these design
decisions become more obvious. Most notably, sending non-fixed-width data structures such as 
\emph{hash tables}, user-defined {\em structs} or {\em objects}, or even \emph{strings} is a challenging task.

We overcome this inconvenience by serializing structured message contents into variable-length byte arrays.
\texttt{YGM}'s message buffering strategy concatenates and transmits these byte arrays as conventional MPI messages and destination
nodes deserialize the arrays back into their original data structure forms.
This pattern affords the transparent communication of structured data with no additional complexity imposed on the application, 
at the cost of a small amount of computing overhead in the serialization and deserialization steps.

Serializing messages in this manner provides two main benefits. First, we are able to send variable length
objects, such as strings, without padding. We take advantage of this capability in the experiment of
Sec.~\ref{sec:wdc_fqdn}. Second, we are able to easily send messages with payloads of different
types in arbitrary orders.
This feature relies upon the \texttt{cereal} C++ serialization library in its implementation \cite{cereal}.

\subsubsection{RPC Semantics}
Messages in \texttt{YGM} have three basic components: a function to execute, arguments to pass to the
function, and an MPI rank at which to evaluate the function. By wrapping the user-provided function in a \texttt{C++} \emph{lambda} function and
serializing the user-provided function arguments along with a pointer to the user function (offsets  to
\texttt{C++} lambdas are handled by senders and receivers to account for address space randomization), \texttt{YGM} can send a collection of serialized bytes through MPI, with the target rank 
unpacking and evaluating the user-provided function with its arguments appropriately upon receipt.

This method of serializing function arguments and wrapping functions mimics some functionality of a \emph{remote
procedure call} (RPC) system, allowing users to interleave several different functions to be executed asynchronously.
One major deviation from traditional RPC programming is that responses are not sent upon completion of requests, giving
\texttt{YGM} a \emph{fire-and-forget} model. This paradigm lends itself naturally to many graph problems,
where a vertex's MPI rank performs a small computation and sends a message to activate some number of its neighbors' ranks
 so they can perform a further calculation.

\subsubsection{Containers}
\label{sec:containers}

The interleaving of messages afforded by \emph{fire-and-forget} RPC semantics makes it possible to build
composable containers on top of \texttt{YGM}'s communication framework to handle many simple tasks that arise frequently in distributed computing. One example that gets
used frequently in this work is a distributed map.
This structure stores \emph{key-value} pairs at deterministic MPI ranks based on a hash of the keys.

One particular use for this distributed map is in building a
distributed \emph{counting set} that keeps individual counts of different items seen across ranks. This structure stores
a small cache on each rank to keep values seen recently, which must be flushed and have its contents sent across the
network occasionally.

This \emph{counting set} is not necessary if merely doing simple triangle counting.
However, for complicated surveys with multiple types of metadata triangles, it is highly useful and we utilize it for the experiments 
in Sec.~\ref{sec:eval}.
Through the interleaving of messages calling different functions, the
counting sets are free to increment counters on remote MPI ranks when their caches are flushed without ever interfering
with the messages used to identify triangles. This composability with pre-built distributed containers allows our system
to answer nontrivial analysis questions involving topology and metadata jointly.

\subsection{Graph Storage}
For triangle identification, \texttt{TriPoll} stores the degree-ordered directed graph $\mathcal{G}_+$, as defined in
Sec.~\ref{sec:preliminaries}, in a custom structure built on \texttt{YGM}'s distributed map container, as described in Sec.~\ref{sec:containers}.

For a vertex $u$, \texttt{TriPoll} stores a unique identifier associated to $u$ as the key. Values are a pair containing $meta(u)$ and an
adjacency list augmented to contain the necessary metadata, $Adj_+^m(u)$, where 
\begin{equation*}
	Adj_+^m(u) = \{ (v, meta(u,v), meta(v)) | v \in Adj_+(u) \}.
\end{equation*}
In $Adj_+^m(u)$, we consider the elements to be ordered by degree. 

We use random or cyclic
partitionings of vertices across MPI ranks and do not attempt to do more sophisticated partitionings in this
work. In large scale-free graphs, high-degree vertices tend to cause the computation and storage imbalances that are one
reason to partition a graph differently. In this case, we construct $\mathcal{G}_+$ to reduce the
degree of these hub vertices. This alleviates the imbalances in triangle identification to a point that cyclic
partitioning becomes palatable. Previous work has shown partitioning to be helpful, specifically if data replication is
allowed \cite{Pandey.2021.TRUST.9373989, zhang2019litete}.

This DODGr structure is designed to be used in a vertex-centric manner. For the high-level algorithms we describe, the
only operations we will need are a way to iterate over all vertices, and a $DODGr.visit(v, func, args)$ operation
that sends an RPC call to $Rank(v)$ to execute the function $func(v, args)$, where passing $v$ to $func$ is understood
to give it access to $v$'s metadata and adjacency information.

The decision to store target vertex metadata along edges changes the storage requirement of vertex metadata
from $\mathcal{O}(| \mathcal{V} |)$ to $\mathcal{O}(| \mathcal{E} |)$.
This additional memory overhead is required because a user-defined callback must have access to all metadata associated with $p, q, r$ and the 3 edges between them once $\Delta_{pqr}$ is identified.
The ordering of $\mathcal{G}_+$ allows us to enumerate $\Delta_{pqr}$ without visiting $r$, the highest-degree vertex involved. 
Including $meta(r)$ in $Adj_+(q)$ maintains this communication advantage at the cost of some additional memory. 
It is also possible to aggregate all target metadata within an MPI rank to
avoid storing redundant copies of $meta(r)$. This, however, does not avoid the worst case storage requirement of
$\mathcal{O}(| \mathcal{E} |)$ for vertex metadata.

\subsection{\emph{Push-Only} Algorithm}
\label{sec:push-only}

\begin{figure}[t]
	\begin{center}
		\includegraphics[width=3.0in]{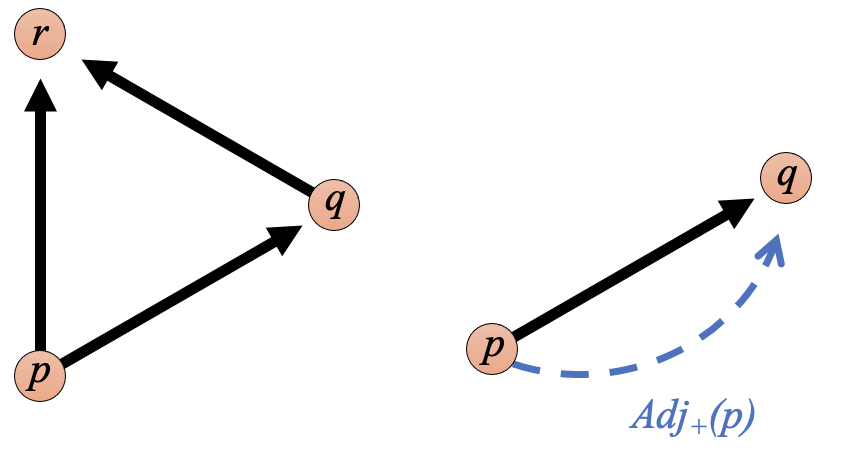} 
		\captionsetup{size=small}
		\caption{Roles of vertices $p <_+ q <_+ r$ in each triangle within the degree-ordered directed graph (DODGr) $\mathcal{G}_+$ (left) and a push message sent from $i$ to $j$ containing the neighborhood of $i$ (right).}
		\label{fig:push}
	\end{center}
\end{figure}

For each vertex $p$, the construction of $\mathcal{G}_+$ guarantees that $p <_+ q$ for all $q \in Adj_+(p)$. 
Vertex $p$ can be a part of $\mathcal{O}(d_+(p)^2)$ triangles, one for each pair of vertices in $Adj_+(p)$. For the
purposes of triangle identification, we can imagine the
process of spawning these $\mathcal{O}(d_+(p)^2)$ wedge checks as iterating through $Adj_+(p)$ in increasing order, popping the
vertex $q$ currently at the front, and sending the remaining adjacency list ($Adj_+(p) \setminus \{v \, : \, v <_+ q
\}$) to $Rank(q)$ as potential vertices (the `$r$' vertices) to search for in
the adjacency list $Adj_+(q)$. This is depicted on the right of Fig.~\ref{fig:push}. 

Additionally, $Rank(p)$ must send $meta(p), meta(p, q),$ and \\$meta(p, r)$ for every $r$ to $Rank(q)$ to be used in a
callback after a triangle $\Delta_{pqr}$ is found. Ultimately, the communication for triangle identification just
described occurs using $Adj_+^m(p)$, rather than $Adj_+(p)$. It is worth noting our definition of $Adj_+^m(p)$ includes
storing $meta(r)$ for all $r \in Adj_+(p)$, which is not needed by $Rank(q)$, as $Rank(q)$ will already have a copy of
$meta(r)$ if $\Delta_{pqr}$ exists. In reality, this extra metadata is never actually transmitted, but for ease of
discussion, we ignore the fact that $Rank(q)$ receives a subset of the metadata found in $Adj_+^m(p)$.

When $Rank(q)$ receives a collection of vertices to search for, we can consider these as a batch of wedge checks, reducing the
number of checks that must be performed. 
We can identify the common elements between the sorted lists $Adj_+^m(q)$ and the subset of $Adj_+^m(p)$ sent to $q$ by traversing each list simultaneously, a process known as a
\emph{merge-path} intersection \cite{green2014fast}.

As triangles are identified using this intersection technique, the callback provided by the user is executed using the
metadata of all vertices and edges in each triangle. As a check, $Rank(p)$ sent $meta(p), meta(p,q)$, and $meta(p,r)$,
to $Rank(q)$, and $Rank(q)$ already had $meta(q)$ and $meta(q,r)$, as well as $meta(r)$ by our construction of
$Adj_+^m(q)$. Therefore, all of the necessary metadata is colocated at $Rank(q)$ at this moment in time, allowing the
callback to execute correctly.

Pseudocode for this algorithm is given in Alg.~\ref{algorithm:distributed_triangle_survey_push_only}. The triangle
survey takes in arguments that are a graph, a callback operation defined by the user to be executed on the metadata of
every triangle, and optional user-provided arguments that will be passed to the callback. These extra arguments often
have their state modified by the user's callback to capture the results of the desired survey.

We refer to this simplistic vertex-centric merge-path
based triangle identification algorithm as the \emph{Push-Only} implementation of \texttt{TriPoll}, for reasons we will discuss in the next section. 

\begin{algorithm}
	\small
	\caption{TriPoll (Push-only)}
	\label{algorithm:distributed_triangle_survey_push_only}
	\begin{algorithmic}[1]
		\Procedure{Triangle\_Survey}{$\mathcal{G}, user\_callback, user\_args$}
		\State $\mathcal{G}_+ \leftarrow DODGr(\mathcal{G})$
		\Comment{described in Sec.~\ref{sec:preliminaries}}
		\ForAll {$p \in \mathcal{G}_+$}
		\ForAll {$q \in Adj_+(p)$}
		\State $\mathcal{G}_+.visit(q, \lambda\_intersection, Adj_+^m(p))$
		\EndFor
		\EndFor
		\State \textbf{barrier}
		\EndProcedure
		\State

		\algrenewcommand\algorithmicfunction{}
		\Function{$\lambda\_intersection$}{$q\in\mathcal{G}_+, Adj_+^m(p))$}
		\Comment{invk. on $q$, recv. $Adj_+^m(p)$}
		\ForAll {$r \in Adj_+(p) \cap Adj_+(p)$}
		\Comment{merge-path intersection}
		\State $user\_callback(meta(\Delta_{pqr}), user\_args)$
		\EndFor
		\EndFunction
	\end{algorithmic}
\end{algorithm}

\subsection{\emph{Push-Pull} Optimization}

\begin{figure}[t]
	\begin{center}
		\includegraphics[width=2.3in]{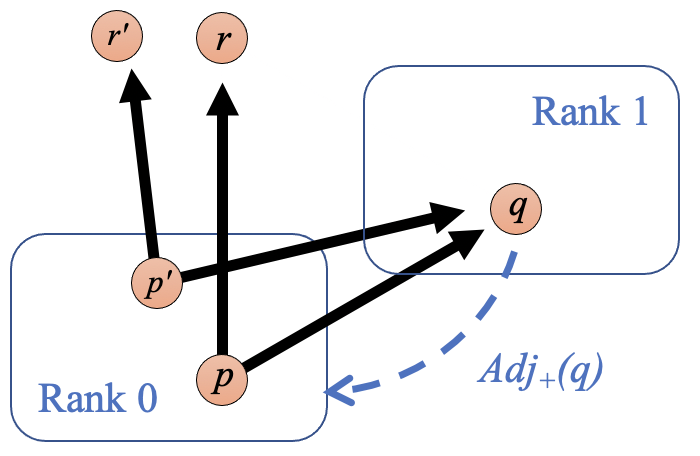} 
		\captionsetup{size=small}
		\caption{A situation depicting when the pull approach is useful.  Assume $p, p'$ are stored on $Rank(p)$ and
		$|Adj_+(p) \setminus \{v \, : \, v <_+ q \} | + |Adj_+(p') \setminus \{v \, : \, v <_+ q \} | >>  |Adj_+(q)|$.
	Then it is potentially advantageous to send $Adj_+(q)$ to $Rank(p)$ so that communication is minimized.}
		\label{fig:pull}
	\end{center}
\end{figure}

Given the $\mathcal{O}(d_+(p)^2)$ wedge checks generated by every vertex, each of which involves checking for the
existence of an edge potentially stored on another compute node, distributed triangle identification spawns massive amounts of
network traffic. Any reduction in this volume of communication has the potential to greatly speed up computations.

The process we described in Section \ref{sec:push-only}, which we refer to as \emph{pushing}, involves $Rank(p)$
identifying a vertex $q$ and a list of candidate $r$ vertices, and then ``\emph{pushing}'' these candidates to $Rank(q)$ to identify vertices that complete
triangles. Alternatively, we can envision $Rank(p)$ identifying vertex $q$ and asking $Rank(q)$ to send the adjacency
list $Adj_+^m(q)$ for $Rank(p)$ to enumerate triangles, instead ``\emph{pulling}'' vertex $q$'s adjacency list. 

The addition of a pull capability is beneficial if $Rank(p)$ knows $d_+(q)$ in advance, which requires only a small
constant amount of additional memory per edge.
However, its true advantage occurs when pulls are coalesced within a computational process, resulting in each MPI rank pulling
$Adj_+^m(q)$ at most once, rather than potentially once per vertex stored locally. For each
process $Proc$, if
$$
|Adj_+(q)| <  \sum_{p' \in Proc} \left| Adj_+(p') \setminus \{v \, : \, v <_+ q \}   \right|,
$$
the pulling approach likely saves on total communication.   See Fig.~\ref{fig:pull} for a simple illustration.
Note that sending metadata for use when triangles are detected 
makes the savings difficult to estimate precisely a-priori.

In the \emph{Push-Pull} \texttt{TriPoll} algorithm, we choose whether to push from $Rank(p)$ or pull from $Rank(q)$ for each target
vertex by taking an initial conditional pass over the data. 
This pass computes the total number of edges that will be sent to each target vertex summed across all local vertices, but does not
actually transmit any adjacency information. 
The pass also stores pointers to efficiently iterate over source vertices stored locally, preparing for the eventuality
that the target's adjacency list is pulled.

After this communication-free counting is done, each MPI rank sends a single message to each processor responsible for one of its target vertices $q$
with the total number of edges it intends to send. 
Then $Rank(q)$ determines whether pulling would result in less communication than pushing. 
Correspondingly, $Rank(q)$ either adds the source rank to a list of ranks to which to send $Adj_+^m(q)$, or sends a reply
informing the source rank that the proposed pull is ill-advised. Due to this determination of whether remote vertices should be
pulled or not mimicking the \emph{Push-Only} algorithm's pass over adjacency lists, but without sending adjacency information,
we call this the \emph{Push vs Pull Dry-Run}.

After these determinations are made, each rank iterates over its local vertices in a \emph{Push Phase}: for each $p$, it sends a subset of
$Adj_+^m(p)$ to vertex $q$ for all $(p, q)$ such that $Adj_+^m(q)$ is not set to be pulled. Finally, each rank iterates over
the list of ranks each of its local vertices is being pulled from to complete the sensible pulls, coalesced where possible, in what we
call the \emph{Pull Phase}.

\subsection{Callback Examples}

In Alg.~\ref{algorithm:distributed_triangle_survey_push_only}, we see there is nothing returned by \texttt{TriPoll}.
Instead, we rely on the user-defined callback that gets executed on Line 11 to produce the side-effects desired to
function as an output.

The simplest example of a callback is incrementing a counter to perform triangle counting. This is shown in
Alg.~\ref{algorithm:triangle_counting}. In this case, the callback is given all of the metadata associated to a
triangle, which it completely ignores. At the end of this survey, each MPI rank has a local count of triangles seen that
must be combined in an $All\_Reduce$-type operation.

\begin{algorithm}
	\small
	\caption{Triangle Counting}
	\label{algorithm:triangle_counting}
	\begin{algorithmic}[1]
		\Procedure{Triangle\_Count}{$\mathcal{G}$}
		\State $tc \leftarrow 0$
		\State $Triangle\_Survey(G, \lambda\_triangle\_count, tc)$
		\State $global\_tc \leftarrow all\_reduce(tc, SUM)$
		\Comment{triangle count}
		\State\Return $global\_tc$
		\EndProcedure
		\State

		\algrenewcommand\algorithmicfunction{}
		\Function{$\lambda\_triangle\_count$}{$meta(\Delta_{pqr}), tc$}
		\State $tc \leftarrow tc+1$
		\EndFunction
	\end{algorithmic}
\end{algorithm}

As a slightly more involved example, suppose we wish to know the distribution of maximum edge labels seen among all
triangles in which all vertex labels are distinct. This data can be gathered using
Alg.~\ref{algorithm:max_edge_example}, where we have made use of the \emph{distributed counting set} described in
Sec.~\ref{sec:containers}.

\begin{algorithm}
	\small
	\caption{Max Edge Label Distribution}
	\label{algorithm:max_edge_example}
	\begin{algorithmic}[1]
		\Procedure{Max\_Edge\_Label\_Distribution}{$\mathcal{G}$}
		\State $counters \leftarrow distributed\_counting\_set$
		\State $Triangle\_Survey(G, \lambda\_max\_edge\_counts, counters)$
		\State\Return $counters$
		\EndProcedure
		\State

		\algrenewcommand\algorithmicfunction{}
		\Function{$\lambda\_max\_edge\_count$}{$meta(\Delta_{pqr}), counters$}
		\If{$meta(p) \neq meta(q) \neq meta(r)$}
		\State $max\_edge \leftarrow max(meta(pq), meta(pr), meta(qr))$
		\State $counters.increment(max\_edge)$
		\EndIf
		\EndFunction
	\end{algorithmic}
\end{algorithm}

At the conclusion of Alg.~\ref{algorithm:max_edge_example}, the returned $counters$ object contains the result of
this particular triangle survey. At that point, the user may use this data for any desired visualization purposes or as
part of some further processing.

\section{Evaluation}
\label{sec:eval}

Here we present an evaluation of the performance of \texttt{TriPoll}, as well as the results of the surveys.
We begin with strong and weak scaling experiments in Sec~\ref{sec:strong_scaling} and \ref{sec:weak_scaling}, followed
by comparisons to related work in Sec~\ref{section:related_work_empirical_comparison}. These studies are all done in
the case of simple triangle counting, a subset of the functionality \texttt{TriPoll} provides. We then show the results of
experiments on a large Reddit dataset with timestamps to gather distributions of
the time required for triangles to fully form (which we refer to as {\em closure rate}) in
Sec.~\ref{sec:reddit_experiment} and provide the strong scaling behavior. Following this, we give results of another
experiment designed to study the domain name metadata present in the 224 billion edge Web Data Commons 2012 graph. We finish by looking at the performance impact
of including nontrivial metadata (Sec.~\ref{sec:metadata_performance}) and the effects of the \emph{Push-Pull}
optimization (Sec.~\ref{sec:push-pull_performance}) on synthetic and real datasets.

\subsection{Test System}
Experiments presented here are performed on the \emph{Catalyst} cluster at LLNL with nodes featuring dual Intel Xeon E5-2695v2 processors
totaling 24 cores along with 128GB of DRAM per node. \emph{Catalyst} uses an Infiniband QDR interconnect.


\subsection{Datasets}
\label{sec:datasets}
For the strong scaling experiments and comparison to related work, we use several openly available datasets \cite{friendster, Kwak10www, Boldi2011uk-web-07, meusel2014webcc,
WDC2012}, most of which are available from \cite{Rossi2015network_repository}. For weak scaling experiments, we use
R-MAT graphs \cite{chakrabarti2004rmat} up to scale 32.

We downloaded the Reddit dataset from \cite{pushshift}, and the portion we use contains all comments and posts scraped from Reddit in the time period of December 2005
to April 2020. This amounts to 835 million comment authors and 7.2 billion unique comments. The graph we construct uses authors as vertices and comments between authors as undirected edges. This data is naturally
given in the form of a multigraph with authors possibly replying to each others' comments multiple times. The graph we
use keeps the chronologically-first comment between two authors and discards the remaining comments. This reduces the edge count
to 9.4 billion edges.

Tab.~\ref{tab:datasets} gives an overview of the non-synthetic graphs used. All graphs were treated as
undirected. For consistency, we report edge counts as the number of directed edges in a graph after
symmetrizing or, equivalently, the number of nonzeros in a symmetrized graph's adjacency matrix.

\begin{table}
	\footnotesize
	\begin{tabular}[h]{l c c c c c}
		\toprule
		Graph & $| \mathcal{V} |$ & $| \mathcal{E} |$ & $| \mathcal{T} |$ & $d^{max}$ & $d_+^{max}$ \\
		\midrule
		LiveJournal \cite{backstrom2006livejournal} & 4.85M & 69.0M & 286M & 20333 & 686 \\
		Friendster \cite{friendster} & 66M & 3.6B & 4.2B & 5214 & 868 \\
		Twitter \cite{Kwak10www} & 42M & 2.4B & 34.8B & 3.0M & 4102 \\
		uk-2007-05 \cite{Boldi2011uk-web-07} & 106M & 6.6B & 286.7B & 975K & 5704 \\
		web-cc12-hostgraph \cite{meusel2014webcc} & 101M & 3.8B & 415B & 3.0M & 10654 \\
		Web Data Commons 2012 \cite{WDC2012} & 3.56B & 224.5B & 9.65T & 95M & 10683 \\
		Reddit & 835M & 9.4B & 88.1B & 1.70M & 3301 \\
		\bottomrule
	\end{tabular}
	\captionsetup{size=small}
	\caption{Datasets used for experiments}
	\label{tab:datasets}
\end{table}

\subsection{Triangle Counting with \texttt{TriPoll}}
\label{sec:dummy_metadata}
Triangle counting is the simplest example using \texttt{TriPoll}, in which the callback for when a triangle is identified only
increments a counter. This, however, is the extent of functionality present in most distributed triangle processing
solutions. Exceptions are distributed versions of computing truss decompositions \cite{cohen2008truss},  
where counts of triangles are desired at edges, and computing clustering coefficient 
where local counts of triangles are desired at vertices.
Callbacks designed for these local participation counts would merely increment local counters.
On the other hand, \cite{reza2020indexing} represents an example where counts of metadata triangles (involving vertex labels) 
are needed at vertices or edges for use in accelerating general pattern matching code.

In order to count simple triangles globally or locally in our highly generalizable system, we must affix dummy metadata to all vertices and edges of the graph in question. We use
booleans for this purpose. This addition adds a slight overhead to our storage requirements for a graph as well as the
size of every adjacency list sent across the network.

The pseudocode to perform triangle counting with \texttt{TriPoll} that ignores metadata is given in Alg.~\ref{algorithm:triangle_counting}.

\subsection{Triangle Counting Strong Scaling Results}
\label{sec:strong_scaling}

\begin{figure}[t]
	\begin{center}
		\includegraphics[width=\columnwidth]{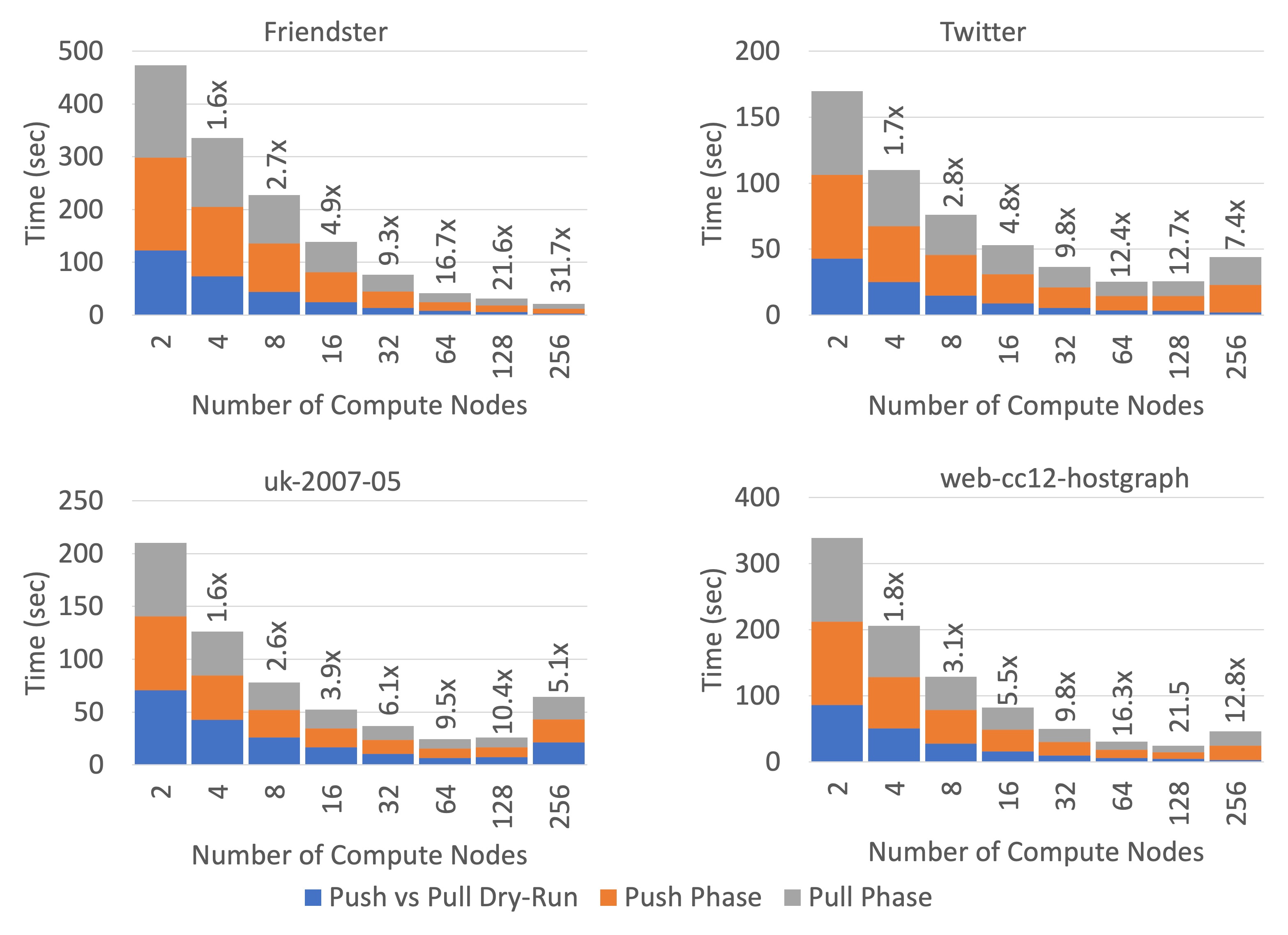}
		\captionsetup{size=small}
		\caption{Strong scaling of each phase of \emph{Push-Pull} algorithm on 2 to 256 compute nodes. Numbers above
		bars indicate the overall speedup on each dataset relative to times using 2 nodes.}
		\label{fig:strong_scaling}
	\end{center}
\end{figure}

We provide a collection of strong scaling studies using \texttt{TriPoll}. For this endeavor, we 
use the openly available topology-only Friendster, Twitter, uk-2007-05, and web-cc12-hostgraph graphs\cite{friendster, Kwak10www, Boldi2011uk-web-07, meusel2014webcc}. 
The results of these triangle counting studies for each of these datasets are given in
Fig.~\ref{fig:strong_scaling}.

On these datasets, we find the \emph{Push-Pull} implementation of \texttt{TriPoll} scales well to between 64 and 256 compute nodes, depending on the dataset. By
the nature of our \emph{Push-Pull} algorithm, we do not expect to see ideal speedup in a strong scaling experiment. As
the number of MPI ranks increases with a fixed graph size, the number of edges allocated to each rank decreases. With
this decrease in edges per rank comes fewer opportunities for the aggregation of edges we exploit in our
\emph{Push-Pull} optimization.

For all datasets except Friendster, running times for \texttt{TriPoll} increase between the 128 and 256 compute node configurations. This is
likely due to limitations in the current implementation of our communication library. At 256 compute nodes, we have
6144 MPI ranks that may be communicating with each other, giving $\binom{6144}{2} \approx 18.8$ million pairs of ranks
potentially communicating. This likely leads to a significant number of small messages being sent because of a lack of
opportunities for message aggregation. This can likely be remedied by adding extra aggregation of messages at the level
of compute nodes (instead of just individual MPI ranks), similar to other work \cite{Priest2019Grapl,
maley2019conveyors}.

%
%

\subsection{Weak Scaling in R-MAT Graphs}
\label{sec:weak_scaling}

We performed weak scaling experiments using R-MAT graphs. 
These graphs provide a standard test for scalability despite their relatively low triangle count for their size. Here we
again use triangle counting as our application to test scalability.

\begin{figure}[t]
	\begin{center}
		\includegraphics[width=2.3in]{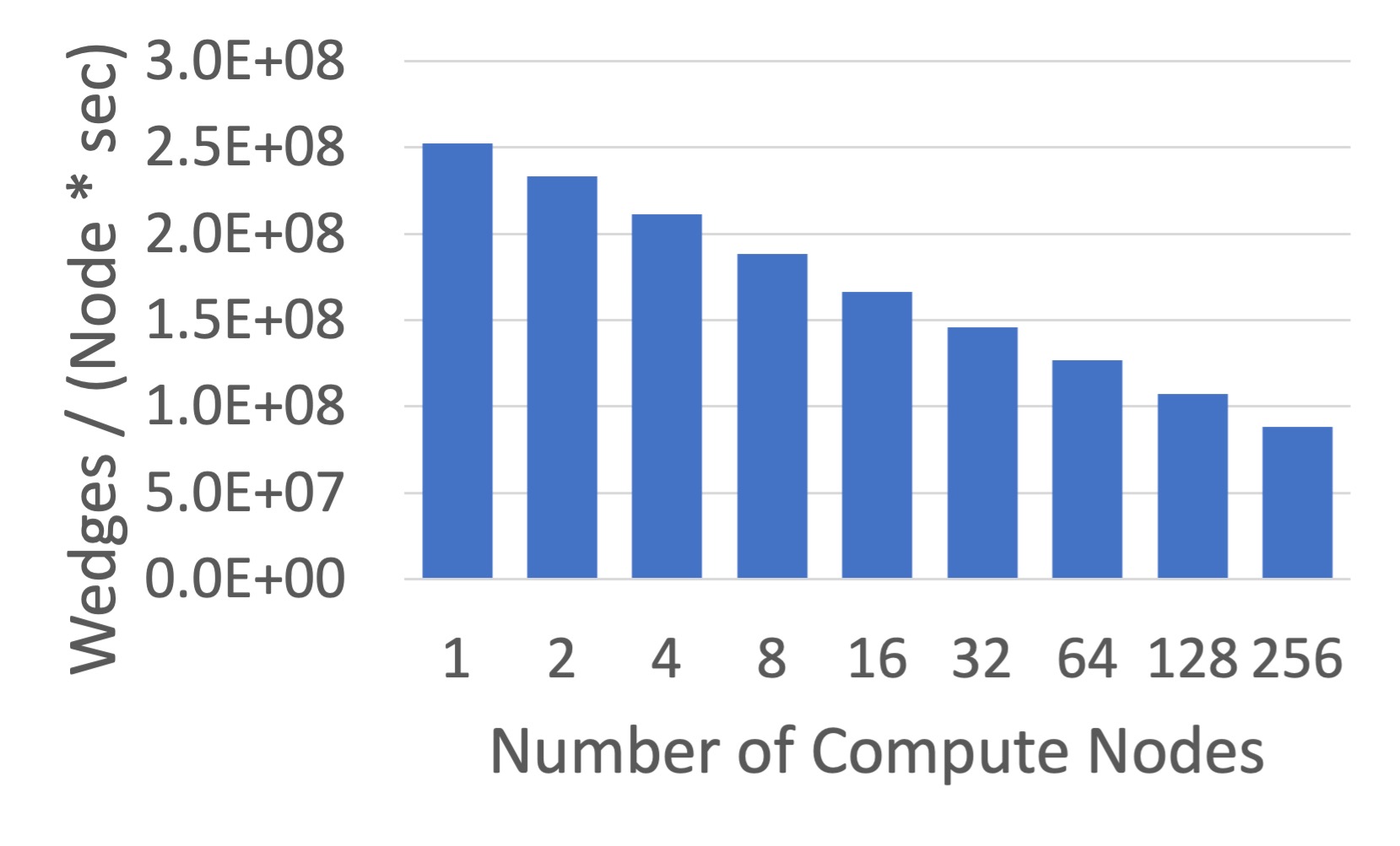}
		\captionsetup{size=small}
		\caption{Weak scaling of triangle counting using \texttt{TriPoll}. Experiments are configured to use a scale 24 R-MAT per compute node, starting with a
	scale 24 on 1 node and reaching up to a scale 32 on 256 nodes.}
		\label{fig:weak_scaling}
	\end{center}
\end{figure}

To assess the weak scalability of our \emph{Push-Pull} algorithm with \texttt{TriPoll}, we want to make sure the rate at which each compute
node is performing work remains roughly constant. Because we have an algorithm whose computational complexity is
$\mathcal{O}(|\mathcal{W}_+|) $ 
(the number of wedges in the DODGr graph $\mathcal{G}_+$) it makes sense to
gauge the weak scalability of a distributed version of the algorithm versus $|\mathcal{W}_+| / (N * t)$ where $N$ is the
number of compute nodes and $t$ is the time required to count triangles. We therefore use this as the vertical axis of
our weak scaling results in Fig.~\ref{fig:weak_scaling}.

We can see that the rate of work per compute node steadily decreases as the number of compute nodes increases. As in the
case of strong scaling, we expect to see this effect, although for a different reason. As the graph grows while keeping
roughly the same number of edges on each compute node, each edge stored has more potential target vertices to connect
to. Likewise, any pair of edges stored on an MPI rank has a decreasing probability of connecting to the same target
vertex as the size of the graph increases. This leaves us with fewer opportunities for the aggregated communication our
algorithm exploits. We will investigate the implications of this effect in Sec.~\ref{sec:push-pull_performance}.

\subsection{Comparison with the Related Work}
\label{section:related_work_empirical_comparison}

Since we are not aware of any system that directly solves the problem addressed by \texttt{TriPoll}, we focus on evaluating
the performance of the core operation of our system, i.e., triangle identification/counting. Using four real-world
graphs, we present empirical comparisons with three tailor-made, MPI-based, distributed triangle counting solutions that
target processing on CPU clusters. Tab.~\ref{table:related_work_comparison} lists end-to-end runtimes of our work, and
solutions by Pearce et al.~\cite{pearce2017triangle}, Tom et
al.~\cite{Tom.2019.ICPP.DistributedTriangle.10.1145/3337821.3337853} and, the 2020
GraphChallenge~\cite{graphchallenge.mit.2017} winner, TriC~\cite{Ghosh.2020.HPEC.Tric.9286167}. We wanted to compare
with~\cite{Acer.2019.HPEC.DistributedTriangle.8916302} which also targets CPU clusters, and demonstrated respectable
performance and scalability; unfortunately the implementation was not publicly available. 

The graphs chosen for these comparisons were the LiveJournal \cite{backstrom2006livejournal}, Friendster \cite{friendster}, Twitter
\cite{Kwak10www}, and Web Data Commons 2012 \cite{WDC2012} graphs. These graphs form a widely-used subset of the
openly available datasets used in benchmarking graph algorithms and provide a wide range of sizes. Details of these
graphs can be found in Tab.~\ref{tab:datasets}

We ran most of these experiments on 1,024 cores (64 compute nodes), except for the Web Data Commons graph. 
This configuration was chosen because the work of Tom et al. requires a number of MPI ranks that is a perfect square. Due to its high
memory demand, we had to run TriC with 256 compute nodes and 4 processes per node to finish triangle counting on
Friendster; for Twitter, experiments with TriC crashed due to insufficient memory, even when using 256 compute nodes.
For the medium-sized graphs, \texttt{TriPoll} outperforms Pearce et al.:
for Friendster, $\sim$1.8$\times$ faster; for Twitter, $\sim$6.8$\times$ faster. 

Although, for the Friendster and Twitter graphs, the work by Tom et al. achieves the fastest
time-to-solution, we were unable to get their code to run with more than 1024
MPI ranks, which confirms its optimizations are geared
towards throughput rather than scalability. In comparison, we demonstrate scalability of \texttt{TriPoll} using up to 224 billion edge graphs and
6144 cores, to the best of our knowledge, the largest scale for real-world graphs to date. The only previous work that
could count triangles in this graph is that of Pearce et al.; however, \texttt{TriPoll} is $\sim$1.8$\times$ faster when
using 256 compute nodes with 24 processes per node. 

To the best of our knowledge, \cite{yasar2020blockbased} demonstrated the second largest scale for
real-world graphs: using the 2014 version of the Web Data Commons graph which has 124B edges (note: the
2014 graph is approximately half the size of the 2012 graph), on an Nvidia DGX machine with
eight GPUs (each with 32GB memory), it counted over four trillion triangles in under two minutes. Although, they offer
superior throughput, other distributed GPU-based solutions (discussed in Sec.~\ref{section:related_work}) did not
demonstrate scalability using real-world graphs as large as we do in this paper; hence, we limit comparison to solutions
that target CPU clusters.     


\begin{table}[!ht]
	\footnotesize
	\renewcommand{\arraystretch}{1.2}
	\setlength{\tabcolsep}{3.0pt}
	\begin{center}
		\begin{tabular}{lrrrr}
			\toprule

			Graph & TriPoll & Pearce et al.~\cite{pearce2017triangle} & Tom et
			al.~\cite{Tom.2019.ICPP.DistributedTriangle.10.1145/3337821.3337853} &
			TriC~\cite{Ghosh.2020.HPEC.Tric.9286167} \\
			\midrule
			LiveJournal & 1.01s & 1.08s & 1.45s & 1.24min\\
			Friendster & 38.62s & 69.79s & 23.78s & 5.55min*\\
			Twitter & 28.96s & 196.10s & 16.43s & N/A\\
			\midrule
			Web Data & 456.7s & 808.7s & N/A & N/A\\
			Commons &  &  &  & \\
			\bottomrule
		\end{tabular}
	\end{center}
	\captionsetup{size=small}
	\caption{End-to-end runtime comparison using four real-world graphs. All experiments were run on
	1024 cores on a 64 node deployment, except for Web Data Commons experiments which use 6144 cores on 256 nodes, and
	the TriC Friendster experiment. *Experiments were run on 256 compute nodes, 4 processes per node.}
	\label{table:related_work_comparison}
\end{table}

\subsection{Triangle Closure Times in Reddit}
\label{sec:reddit_experiment}

\begin{figure}[t]
	\begin{center}
		\includegraphics[width=2.9in]{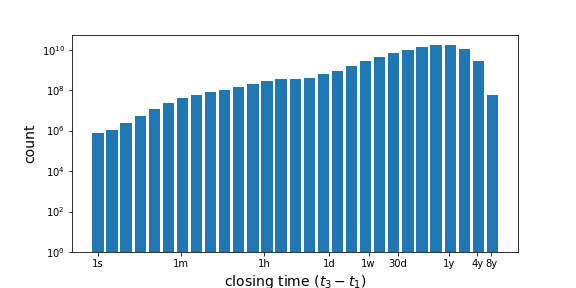} \\
		\includegraphics[width=2.9in]{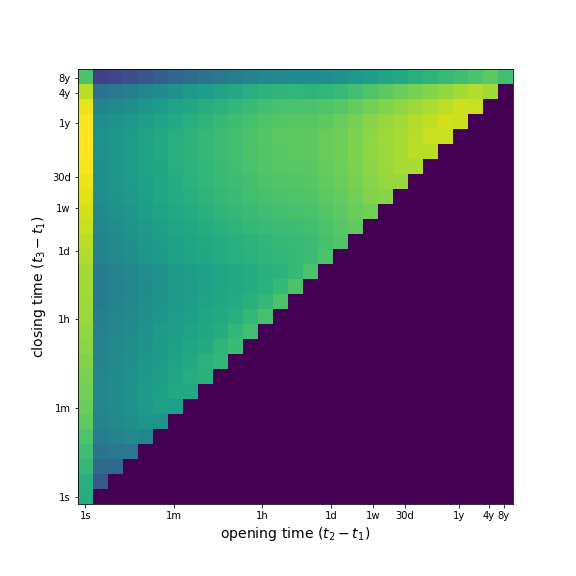}
		\captionsetup{size=small}
		\caption{Distribution of closing time and joint distribution of closing time versus opening time for the Reddit graph. Axes and counts are in log-scales.}
		\label{fig:triangle_time}
	\end{center}
\end{figure}

In our Reddit experiments, we want to determine how quickly the closing edge for each triangle appears relative to the
speed that the initial wedge forms. 
This may be viewed by a network scientist as an interesting dynamic property of triangles, and 
different networks will likely have very different distributions of closure time.   Moreover, within a single network, closure times 
are likely to indicate the nature of certain activity. 
For example, a triangle formed by human peer-2-peer connectivity should happen in reasonable time scales 
(e.g., the time it takes for a human to notice, read, and reply to another's Reddit comment) whereas much faster completion time 
might indicate a coordinated machine activity. 

The graph we construct for these experiments uses authors as vertices and comments between authors as undirected edges.
Details of this 9.4 billion edge graph are given in Sec.~\ref{sec:datasets}.

The question we are answering concerns the time of comments. We store these timestamps as edge metadata and do not
make use of vertex metadata.

Once a triangle is found, the {\em wedge opening time}, $\Delta t_{open} = t_2 - t_1$, and {\em triangle closing time},  $\Delta t_{close} = t_3 - t_1$, are computed, where $t_1 \leq t_2
\leq t_3$ are the timestamps of the three edges involved in a triangle in sorted order. A counter for the pair \\ $(\lceil \log_2(\Delta t_{open}) \rceil,
\lceil \log_2(\Delta t_{close})\rceil)$ is then incremented using a distributed counting data structure. Pseudocode for
this application is given in Alg.~\ref{algorithm:reddit_closure_times}. In this case, our counting set is counting pairs
of objects because we are looking for the joint distribution of $\Delta t_{open}$ and $\Delta t_{close}$.

\begin{algorithm}
	\small
	\caption{Reddit Triangle Closure Times}
	\label{algorithm:reddit_closure_times}
	\begin{algorithmic}[1]
		\Procedure{Reddit\_Triangle\_Time\_Joint\_Distribution}{$\mathcal{G}$}
		\State $counters \leftarrow distributed\_counting\_set$
		\State $Triangle\_Survey(G, \lambda\_triangle\_times, counters)$
		\State\Return $counters$
		\EndProcedure
		\State

		\algrenewcommand\algorithmicfunction{}
		\Function{$\lambda\_triangle\_times$}{$meta(\Delta_{pqr}), counters$}
		\If{$meta(p) \neq meta(q) \neq meta(r)$}
		\State $t\_1 \leftarrow min(meta(pq), meta(pr), meta(qr))$
		\State $t\_2 \leftarrow median(meta(pq), meta(pr), meta(qr))$
		\State $t\_3 \leftarrow max(meta(pq), meta(pr), meta(qr))$
		\State $open\_time \leftarrow ceil(log_2(t\_2 - t\_1))$
		\State $close\_time \leftarrow ceil(log_2(t\_3 - t\_1))$
		\State $counters.increment(pair(open\_time, close\_time))$
		\EndIf
		\EndFunction
	\end{algorithmic}
\end{algorithm}

Fig.~\ref{fig:triangle_time} shows
the distributions of closing and opening times generated by this experiment. In the joint distribution, we see that wedges are often formed quickly, but triangles are not (on average)
systematically closed rapidly after wedges are formed.    

\begin{figure}[t]
	\begin{center}
		\includegraphics[width=2.4in]{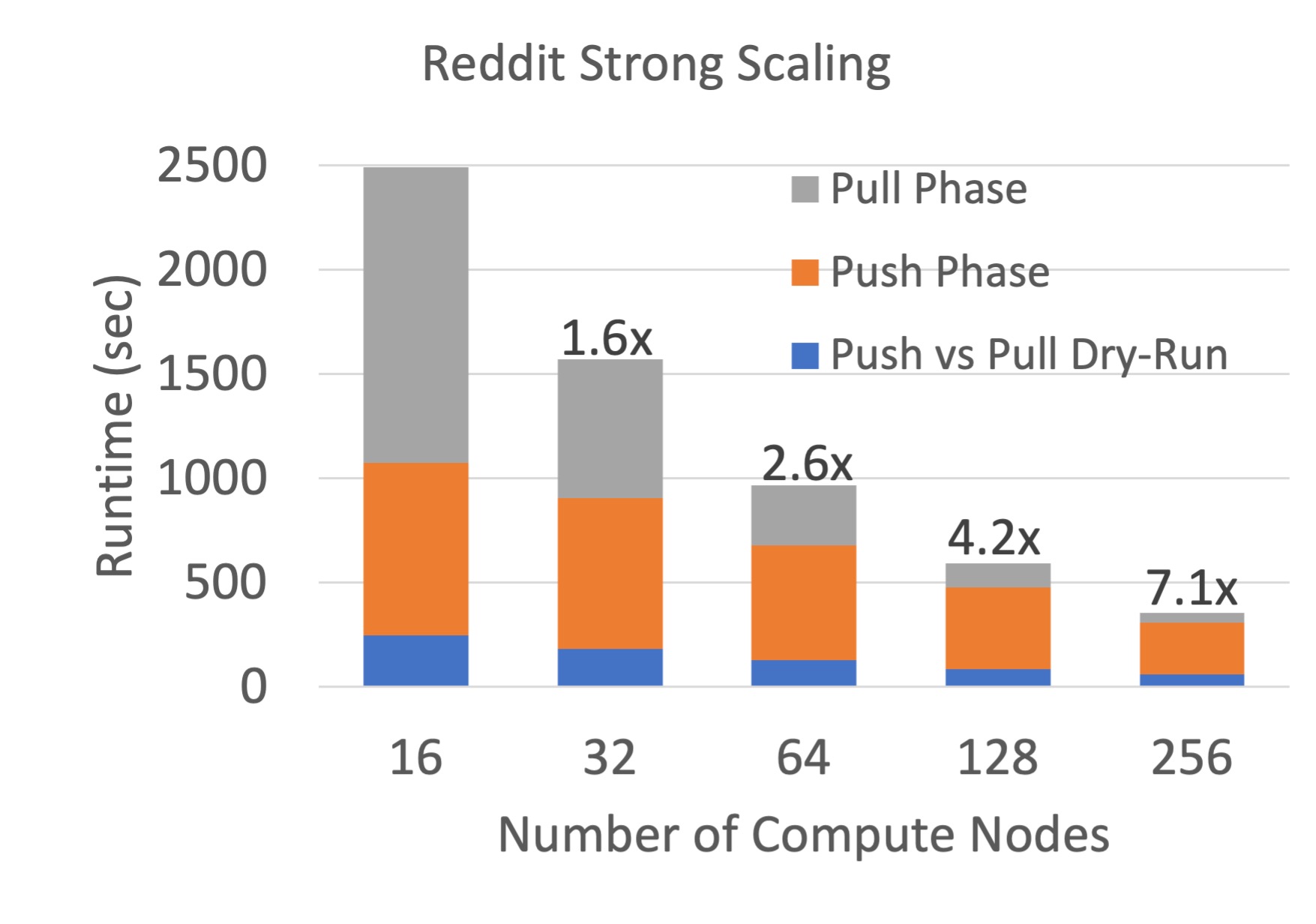}
		\captionsetup{size=small}
		\caption{Strong scaling of triangle closure time collection on up to 256 nodes using \emph{Push-Pull}
			algorithm. Times are divided into 3 pieces:
		the time to determine which vertices to pull, the push phase, and the pull phase. Numbers shown above bars are the
		overall speedup relative to 16 nodes.}
		\label{fig:reddit_scaling}
	\end{center}
\end{figure}

Fig.~\ref{fig:reddit_scaling} shows the strong scaling performance of the \emph{Push-Pull} implementation of \texttt{TriPoll} out to 256 compute
nodes. Performance scales well out to 256 compute nodes for this problem. We do not appear to witness the same issues
with strong scaling discussed in
Sec.~\ref{sec:strong_scaling}. This is likely due to the topology of this graph, as we see the same scaling behavior
from Friendster, another large social network graph.

Within Fig.~\ref{fig:reddit_scaling}, it is important to note that the breakdowns of times into the different phases of
computation does not necessarily show our pull phase scaling well and the push phase showing limited scalability. These results
are actually indicative of a shift in the \emph{Push-Pull} algorithm from pulling many vertices when a small number of
nodes is used, to
an almost entirely push-based algorithm on larger allocations. This is due to the decreased opportunities for the
aggregation necessary in our \emph{Push-Pull} optimization when each MPI rank has fewer edges. This can be seen by the decrease
in average vertices pulled per rank shown in Tab.~\ref{tab:reddit_avg_pulls}.

\begin{table}[t]
	\footnotesize
	\begin{center}
	\begin{tabular}{|c|c|}
		\hline
		Nodes & Avg. Pulls Per Rank \\
		\hline
		16 & 861K\\
		\hline
		32 & 466K \\
		\hline
		64 & 228K \\
		\hline
		128 & 101K \\
		\hline
		256 & 42.2K \\
		\hline
	\end{tabular}
	\captionsetup{size=small}
	\caption{Average number of vertices pulled per rank as number of compute nodes increases.}
	\label{tab:reddit_avg_pulls}
\end{center}
\end{table}

\subsection{FQDN Analysis on Web Data Commons 2012}
\label{sec:wdc_fqdn}
We have seen in Sec.~\ref{section:related_work_empirical_comparison} that the work of \cite{pearce2017triangle}
provides the only previous solution which is capable of counting triangles in the Web Data Commons 2012 graph, although
\texttt{TriPoll} outperforms the former handily. Next, we look to perform
an analysis of the triangles found in this webgraph, a capability not possible within the confines of
\cite{pearce2017triangle}.

The vertices of the Web Data Commons graph represent webpages, each with a URL. For this experiment, we extract the
\emph{fully qualified domain name} (FQDN) from each vertex's URL and attach it as vertex metadata.
We store this vertex metadata as \texttt{C++} strings. We are able to do so without padding the strings to a fixed
size or making use of lookup tables to retrieve the FQDNs by leveraging the serialization features described in
Sec.~\ref{sec:comm}. This deliberate choice of using strings prepares us for situations where metadata can truly be arbitrary in
length.

While identifying triangles in this graph, we use a callback that counts 3-tuples of FQDNs involved in triangles, only
counting triangles with 3
distinct FQDNs. This triangle analysis completes in 1694.6, compared to the 456.7 seconds required to count triangles
without vertex metadata included.

During this distributed computation, \texttt{TriPoll} identifies 248.7 billion triangles where all FQDNs are distinct with a
total of 39.2 billion unique 3-tuples of FQDNs. After this, we
post-process the results on a single machine to investigate these FQDN triangles.

\begin{figure}[t]
	\begin{center}
		\includegraphics[width=2.4in]{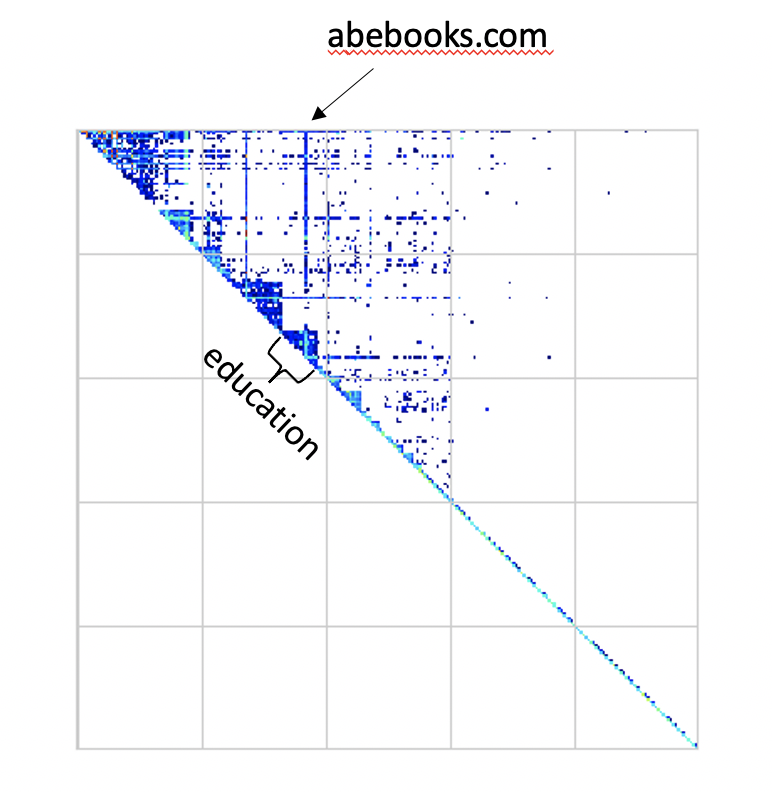}
		\captionsetup{size=small}
		\caption{Distribution of FQDNs involved in triangles with "amazon.com".}
		\label{fig:amazon_fqdn_triangles}
	\end{center}
\end{figure}

As an example, we searched the 39.2 billion 3-tuples of FQDNs in triangles to find all triangles involving
"amazon.com". This generates a 2D distribution of pairs of FQDNs. This distribution is shown in
Figure~\ref{fig:amazon_fqdn_triangles}, where the FQDNs are ordered based on communities identified by the Louvain method.

Within this distribution, we can identify several points of interest. First, there are several relatively dense rows
near the top of Figure~\ref{fig:amazon_fqdn_triangles} corresponding to other Amazon domains and products such as
"amazon.co.uk", "amazon.ca", and "audible.com". Additionally, we have labeled another relatively dense row/column
identified by "abebooks.com", an online retailer competing with Amazon. Unsurprisingly, sites linking to an Amazon
product page are likely to link to the corresponding product at this competing retailer.

We have also identified a community among the FQDNs comprising a large number of domains for educational
institutions and libraries. This community also features a number of booksellers, including abebooks.com. 

This example demonstrates the data we are able to collect and study concerning metadata on triangles in graphs large
enough that few previous works could even perform global triangle counts.

\subsection{Impact of Metadata on Performance}
\label{sec:metadata_performance}

\begin{table*}
	\footnotesize
	\begin{center}
	\begin{tabular}[t]{|l|c|c|c|c|c|c|c|c|}
		\hline
		Dataset & \multicolumn{2}{c|}{Measurement} & 8 Nodes & 16 Nodes & 32 Nodes & 64 Nodes & 128 Nodes & 256 Nodes \\
		\hline
		\multirow{4}{*}{Friendster} & \multirow{2}{*}{Communication Volume (GB)} & Push-Only & 727.7 & 729.6 & 730.6 &
		731.1 & 731.3 & 731.4 \\
									& & Push-Pull & 572.1 & 659.0 & 713.4 & 744.1 & 760.4 & 768.8 \\
									\cline{2-9}
									& \multirow{2}{*}{Runtime (sec)} & Push-Only & 164.9 & 80.4 & 38.1 & 18.6 & 13.8 &
									11.8 \\
									& & Push-Pull & 179.6 & 99.9 & 52.8 & 29.5 & 22.8 & 15.5 \\
		\hline
		\multirow{4}{*}{Twitter} & \multirow{2}{*}{Communication Volume (GB)} & Push-Only & 1283 & 1286 & 1288 & 1289 &
								 1289 & 1289 \\
								 & & Push-Pull & 372.9 & 525.8 & 684.6 & 829.8 & 945.2 & 1025.8 \\
								 \cline{2-9}
								 & \multirow{2}{*}{Runtime (sec)} & Push-Only & 201.0 & 99.2 & 49.5 & 25.7 & 14.5 & 25.1 \\
								 & & Push-Pull & 100.9 & 60.5 & 36.8 & 23.0 & 22.4 & 38.4 \\
		\hline
		\multirow{4}{*}{uk-2007-05} & \multirow{2}{*}{Communication Volume (GB)} & Push-Only & 3323 & 3331 & 3335 & 3337
									& 3338 & 3338 \\
									& & Push-Pull & 625.9 & 943.2 & 1343 & 1737 & 2068 & 2281 \\
									\cline{2-9}
									& \multirow{2}{*}{Runtime (sec)} & Push-Only & 388.8 & 190.7 & 97.1 & 49.1 & 28.0 &
									37.0 \\
									& & Push-Pull & 137.9 & 89.8 & 57.9 & 37.1 & 33.8 & 69.7 \\
		\hline
		\multirow{4}{*}{web-cc12-hostgraph} & \multirow{2}{*}{Communication Volume (GB)} & Push-Only & 18497 & 18545 &
		18569 & 18582 & 18588 & 18591 \\
			  & & Push-Pull & 437.1 & 580.8 & 739.6 & 931.3 & 1197 & 1602 \\
									\cline{2-9}
									& \multirow{2}{*}{Runtime (sec)} & Push-Only & 1618.5 & 858.4 & 452.8 & 269.9 &
									159.7 & 163.1 \\
									& & Push-Pull & 198.2 & 112.0 & 63.5 & 38.1 & 28.7 & 48.3 \\
		\hline
	\end{tabular}
	\captionsetup{size=small}
	\caption{Strong scaling results for \emph{Push-Only} and \emph{Push-Pull} implementations including communication
	costs associated to each.}
	\label{tab:counting_strong_scaling}
\end{center}
\end{table*}

Next, we investigate the effect including nontrivial metadata in \texttt{TriPoll} has on performance. To see this effect,
we repeat the weak scaling experiments of Sec.~\ref{sec:weak_scaling}, but we add each vertex's degree as a replacement
for the dummy metadata used previously. In this configuration, we add a callback that counts the occurrences of triples
$(\lceil \log_2(d(p)) \rceil , \lceil \log_2(d(q))\rceil , \lceil \log_2(d(r))\rceil )$ across all triangles discovered. 
The callback to do this counting involves a simple
hash and logarithm of the degrees of vertices $p, q$, and $r$. This scenario gives us a simple example with a small amount
of vertex metadata and a nontrivial callback operation to compare with the triangle counting results.

\begin{figure}[t]
	\begin{center}
		\includegraphics[width=3in]{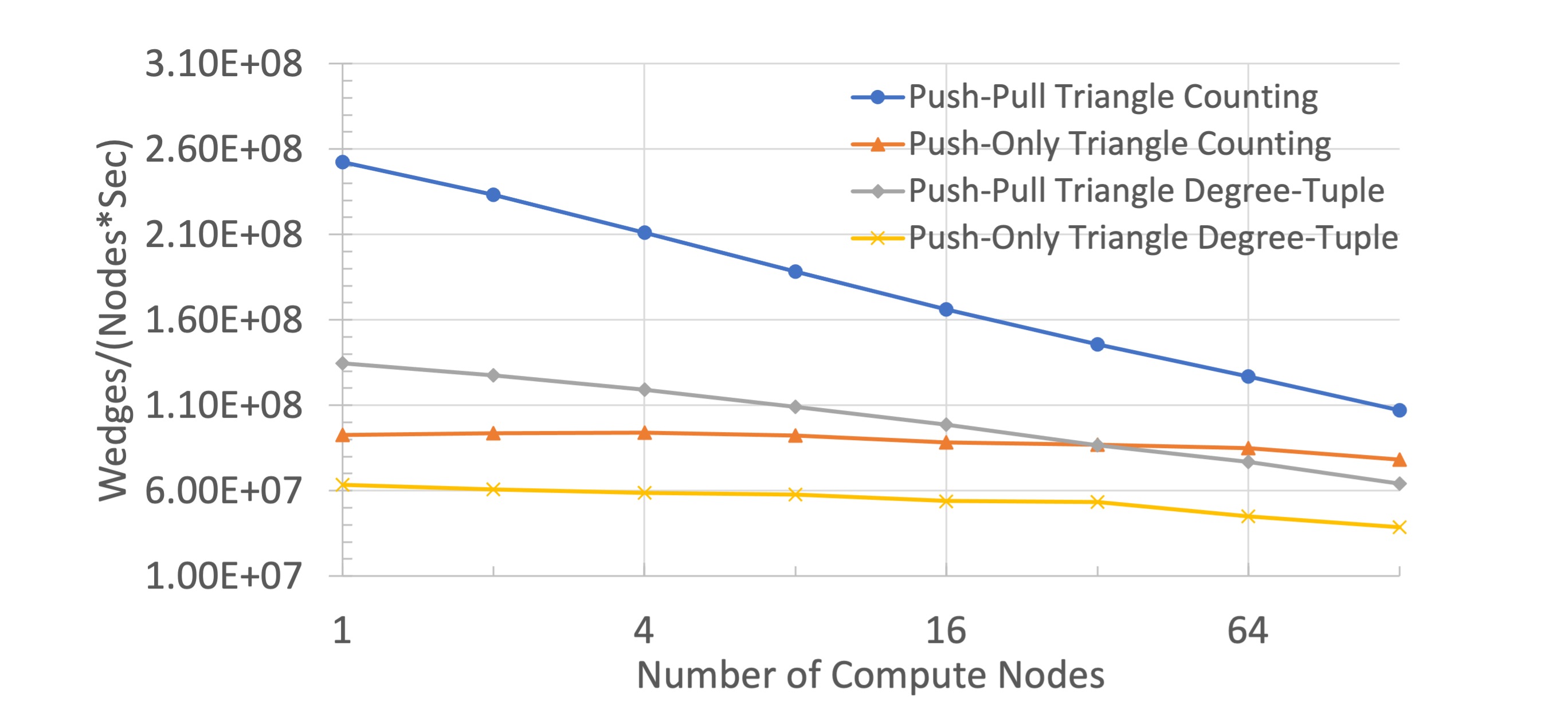}
		\captionsetup{size=small}
		\caption{Effects of metadata inclusion on weak scaling of \emph{Push-Pull} and \emph{Push-Only} algorithms in
		\texttt{TriPoll}}
		\label{fig:metadata_effect}
	\end{center}
\end{figure}

With this experiment we track the performance when using the \emph{Push-Only} as well as
the \emph{Push-Pull} implementations of \texttt{TriPoll}. The results are shown in Fig.~\ref{fig:metadata_effect}.

Each algorithm's throughput is cut
by a factor of just under 2 across all problem sizes as a direct result of the extra metadata and callback. The scalability of
algorithms does not appear to be affected. Although the performance of the \emph{Push-Pull} algorithm
degrades more in absolute terms when metadata is excluded, the relative change in performance remains unchanged
whether or not metadata is included.

\subsection{Push-Pull Optimization Performance}
\label{sec:push-pull_performance}

To study the effectiveness of our \emph{Push-Pull} algorithm over the simpler \emph{Push-Only} algorithm, we repeated
the triangle counting strong scaling experiments of Sec.~\ref{sec:strong_scaling} with the \emph{Push-Only} algorithm. These full
strong-scaling experiments can be found in Tab.~\ref{tab:counting_strong_scaling}. These results include the total volume of data
communicated during the course of each experiment.

These graphs showcase the advantages and disadvantages of the \emph{Push-Pull} implementation. Datasets such as
Friendster provide very little opportunity to reduce communication through pulling adjacency lists. In this scenario,
the overhead of an additional pass over all of the edges of the graph to determine the pulling and pushing behavior is
apparent from the times reported being slower than the \emph{Push-Only} implementation.

This is to be contrasted with the somewhat extreme example provided by the web-cc12-hostgraph, where the communication
volume is reduced by more than a factor of 10, even when using 256 compute nodes (6144 MPI ranks). This massive reduction in network
traffic leads to the \emph{Push-Pull} implementation beating \emph{Push-Only} by a factor of 6 throughout the regime
that both are scaling properly.

Tab.~\ref{tab:counting_strong_scaling} shows a dramatic increase in communication volume for the \emph{Push-Pull}
algorithm when strong scaling on the uk-2007-05 and web-cc12-hostgraph datasets, approaching a factor of 4 increase when
progressing from 8 to 256 nodes. This is a good indication of the lessened opportunities for the aggregation necessary
within an MPI rank to warrant pulling a vertex's adjacency list. On Friendster, we even see the communication volume of
the \emph{Push-Pull} algorithm overtake that of the \emph{Push-Only} algorithm. In this case, the communication required
to check if vertices should be pulled becomes greater than the communication reduction of pulling the beneficial
vertices.

Both algorithms show performance stagnation or regression at 256 compute nodes, however, the negative effect on the
\emph{Push-Only} algorithm is less pronounced than for the \emph{Push-Pull} algorithm. This effect can also be seen in
the weak scaling of Fig.~\ref{fig:metadata_effect}, where the \emph{Push-Only} algorithm shows very little degradation
in work rate as the number of compute nodes increases. Despite the lower scalability, the actual performance of
\emph{Push-Pull} tends to be comparable to or significantly better than that of the \emph{Push-Only} algorithm even when
using thousands of MPI ranks.

\section{Conclusion}

This work demonstrates a novel capability of performing triangle surveys on massive distributed graphs with metadata on
the edges and vertices using \texttt{TriPoll}.
Networks scientists can use such capabilities to better understand the higher-order interactions within their relational datasets, potentially facilitating  
better models for many graph-related machine learning tasks, and better detection of anomalous activities. We have
demonstrated scalability of \texttt{TriPoll} out to thousands of cores on hundreds of compute nodes using graphs with up to
224 billion edges. We showed the capabilities of our system by performing a triangle timing survey on a temporal graph
built from Reddit. Additionally, we compared the performance of \texttt{TriPoll} on the task of triangle counting, a very
simple example of our capabilities, to several works tailored to this problem and were able to outperform them on
several datasets. This includes counting triangles on a 224 billion edge web graph $\sim$1.8$\times$ faster than the only openly available
software able to handle a problem of this size that we are aware of.


\bibliography{fancy_triangles}{}

\begin{thebibliography}{10}

\bibitem{cereal}
cereal.
\newblock https://github.com/USCiLab/cereal.

\bibitem{ygm}
{YGM}.
\newblock https://github.com/LLNL/ygm.

\bibitem{WDC2012}
{Web Data Commons} webgraph.
\newblock { http://webdatacommons.org/hyperlinkgraph/}, 2012.

\bibitem{graphchallenge.mit.2017}
Graph{C}hallenge.
\newblock https://graphchallenge.mit.edu, 2017.

\bibitem{Acer.2019.HPEC.DistributedTriangle.8916302}
S.~{Acer}, A.~{Ya\c{s}ar}, S.~{Rajamanickam}, M.~{Wolf}, and {\"{U}}.~V.
  {Catalyürek}.
\newblock Scalable triangle counting on distributed-memory systems.
\newblock In {\em 2019 IEEE High Performance Extreme Computing Conference
  (HPEC)}, pages 1--5, 2019.

\bibitem{Hasan2018review}
Mohammad Al~Hasan and Vachik~S. Dave.
\newblock Triangle counting in large networks: a review.
\newblock {\em WIREs Data Mining and Knowledge Discovery}, 8(2):e1226, 2018.

\bibitem{Arifuzzaman.2019.TKD.10.1145/3365676}
Shaikh Arifuzzaman, Maleq Khan, and Madhav Marathe.
\newblock Fast parallel algorithms for counting and listing triangles in big
  graphs.
\newblock {\em ACM Trans. Knowl. Discov. Data}, 14(1), December 2019.

\bibitem{backstrom2006livejournal}
Lars Backstrom, Daniel Huttenlocher, Jon Kleinberg, and Xiangyang Lan.
\newblock Group formation in large social networks: Membership, growth, and
  evolution.
\newblock In {\em KDD'06: Proceedings of the 12th ACM SIGKDD international
  conference on Knowledge discovery and data mining}, volume 2006, pages
  44--54, 01 2006.

\bibitem{pushshift}
Jason Baumgartner.
\newblock pushshift.io website.
\newblock { https://pushshift.io}, 2021.

\bibitem{Benson163}
Austin~R. Benson, David~F. Gleich, and Jure Leskovec.
\newblock Higher-order organization of complex networks.
\newblock {\em Science}, 353(6295):163--166, 2016.

\bibitem{Berry2011ToleratingTC}
Jonathan~W. Berry, Bruce Hendrickson, Randall~A. LaViolette, and Cynthia~A.
  Phillips.
\newblock Tolerating the community detection resolution limit with edge
  weighting.
\newblock {\em Physical review. E, Statistical, nonlinear, and soft matter
  physics}, 83 5 Pt 2:056119, 2011.

\bibitem{Boldi2011uk-web-07}
Paolo Boldi, Marco Rosa, Massimo Santini, and Sebastiano Vigna.
\newblock Layered label propagation: A multiresolution coordinate-free ordering
  for compressing social networks.
\newblock In {\em Proceedings of the 20th International Conference on World
  Wide Web}, WWW '11, pages 587--596, New York, NY, USA, 2011. in \textit{WWW}.

\bibitem{chakrabarti2004rmat}
Deepayan Chakrabarti, Yiping Zhan, and Christos Faloutsos.
\newblock R-mat: A recursive model for graph mining.
\newblock In {\em Fourth SIAM International Conference on Data Mining}, April
  2004.

\bibitem{Chen:2018:GET:3190508.3190545}
Hongzhi Chen, Miao Liu, Yunjian Zhao, Xiao Yan, Da~Yan, and James Cheng.
\newblock G-miner: An efficient task-oriented graph mining system.
\newblock In {\em Proceedings of the Thirteenth EuroSys Conference}, EuroSys
  '18, pages 32:1--32:12, New York, NY, USA, 2018. ACM.

\bibitem{cohen2008truss}
Jonathan Cohen.
\newblock Trusses: Cohesive subgraphs for social network analysis.
\newblock {\em National Security Agency Technical Report}, 2008.

\bibitem{cohen2009graph}
Jonathan Cohen.
\newblock Graph twiddling in a mapreduce world.
\newblock {\em Computing in Science \& Engineering}, 11(4):29--41, 2009.

\bibitem{Dathathri.2018.PLDI.Gluon.10.1145/3192366.3192404}
Roshan Dathathri, Gurbinder Gill, Loc Hoang, Hoang-Vu Dang, Alex Brooks, Nikoli
  Dryden, Marc Snir, and Keshav Pingali.
\newblock Gluon: A communication-optimizing substrate for distributed
  heterogeneous graph analytics.
\newblock In {\em Proceedings of the 39th ACM SIGPLAN Conference on Programming
  Language Design and Implementation}, PLDI 2018, pages 752--768, New York, NY,
  USA, 2018. Association for Computing Machinery.

\bibitem{Dave:2016:GIA:2960414.2960416}
Ankur Dave, Alekh Jindal, Li~Erran Li, Reynold Xin, Joseph Gonzalez, and Matei
  Zaharia.
\newblock Graphframes: An integrated api for mixing graph and relational
  queries.
\newblock In {\em Proceedings of the Fourth International Workshop on Graph
  Data Management Experiences and Systems}, GRADES '16, pages 2:1--2:8, New
  York, NY, USA, 2016. ACM.

\bibitem{Dias.2019.SIGMOD.10.1145/3299869.3319875}
Vinicius Dias, Carlos H.~C. Teixeira, Dorgival Guedes, Wagner Meira, and
  Srinivasan Parthasarathy.
\newblock Fractal: A general-purpose graph pattern mining system.
\newblock In {\em Proceedings of the 2019 International Conference on
  Management of Data}, SIGMOD '19, pages 1357--1374, New York, NY, USA, 2019.
  Association for Computing Machinery.

\bibitem{Ghosh.2020.HPEC.Tric.9286167}
S.~{Ghosh} and M.~{Halappanavar}.
\newblock Tric: Distributed-memory triangle counting by exploiting the graph
  structure.
\newblock In {\em 2020 IEEE High Performance Extreme Computing Conference
  (HPEC)}, pages 1--6, 2020.

\bibitem{Giraph.001}
Giraph.
\newblock Giraph.
\newblock http://giraph.apache.org, 2016.

\bibitem{Gonzalez:2012:PDG:2387880.2387883}
Joseph~E. Gonzalez, Yucheng Low, Haijie Gu, Danny Bickson, and Carlos Guestrin.
\newblock Powergraph: Distributed graph-parallel computation on natural graphs.
\newblock In {\em Proceedings of the 10th USENIX Conference on Operating
  Systems Design and Implementation}, OSDI'12, pages 17--30, Berkeley, CA, USA,
  2012. USENIX Association.

\bibitem{Gonzalez:2014:GGP:2685048.2685096}
Joseph~E. Gonzalez, Reynold~S. Xin, Ankur Dave, Daniel Crankshaw, Michael~J.
  Franklin, and Ion Stoica.
\newblock Graphx: Graph processing in a distributed dataflow framework.
\newblock In {\em Proceedings of the 11th USENIX Conference on Operating
  Systems Design and Implementation}, OSDI'14, pages 599--613, Berkeley, CA,
  USA, 2014. USENIX Association.

\bibitem{green2014fast}
Oded Green, Pavan Yalamanchili, and Llu\'{\i}s-Miquel Mungu\'{\i}a.
\newblock Fast triangle counting on the {GPU}.
\newblock In {\em Proceedings of the 4th Workshop on Irregular Applications:
  Architectures and Algorithms}, IA3 '14, pages 1--8. IEEE Press, 2014.

\bibitem{Gurajada:2014:TDS:2588555.2610511}
Sairam Gurajada, Stephan Seufert, Iris Miliaraki, and Martin Theobald.
\newblock Triad: A distributed shared-nothing rdf engine based on asynchronous
  message passing.
\newblock In {\em Proceedings of the 2014 ACM SIGMOD International Conference
  on Management of Data}, SIGMOD '14, pages 289--300, New York, NY, USA, 2014.
  ACM.

\bibitem{Henderson2012RolX}
Keith Henderson, Brian Gallagher, Tina Eliassi-Rad, Hanghang Tong, Sugato Basu,
  Leman Akoglu, Danai Koutra, Christos Faloutsos, and Lei Li.
\newblock {\em RolX: Structural role extraction and mining in large graphs},
  pages 1231--1239.
\newblock Association for Computing Machinery, 2012.

\bibitem{Hoang.2019.HPEC.DistTC.8916438}
L.~{Hoang}, V.~{Jatala}, X.~{Chen}, U.~{Agarwal}, R.~{Dathathri}, G.~{Gill},
  and K.~{Pingali}.
\newblock Disttc: High performance distributed triangle counting.
\newblock In {\em 2019 IEEE High Performance Extreme Computing Conference
  (HPEC)}, pages 1--7, 2019.

\bibitem{Hong:2015:PFD:2807591.2807620}
Sungpack Hong, Siegfried Depner, Thomas Manhardt, Jan Van Der~Lugt, Merijn
  Verstraaten, and Hassan Chafi.
\newblock Pgx.d: A fast distributed graph processing engine.
\newblock In {\em Proceedings of the International Conference for High
  Performance Computing, Networking, Storage and Analysis}, SC '15, pages
  58:1--58:12, New York, NY, USA, 2015. ACM.

\bibitem{hoory2006expander}
Shlomo Hoory, Nathan Linial, and Avi Widgerson.
\newblock Expander graphs and their application.
\newblock {\em Bulletin (New Series) of the American Mathematical Society}, 43,
  10 2006.

\bibitem{Hu.2018.SC.TriCore.8665796}
Y.~{Hu}, H.~{Liu}, and H.~H. {Huang}.
\newblock Tricore: Parallel triangle counting on {GPU}s.
\newblock In {\em SC18: International Conference for High Performance
  Computing, Networking, Storage and Analysis}, pages 171--182, 2018.

\bibitem{Huang.2018.HPEC.FPGATriangle.8547536}
S.~{Huang}, M.~{El-Hadedy}, C.~{Hao}, Q.~{Li}, V.~S. {Mailthody}, K.~{Date},
  J.~{Xiong}, D.~{Chen}, R.~{Nagi}, and W.~{Hwu}.
\newblock Triangle counting and truss decomposition using {FPGA}.
\newblock In {\em 2018 IEEE High Performance extreme Computing Conference
  (HPEC)}, pages 1--7, 2018.

\bibitem{Kanewala.2018.PASC.10.1145/3218176.3218229}
Thejaka~Amila Kanewala, Marcin Zalewski, and Andrew Lumsdaine.
\newblock Distributed, shared-memory parallel triangle counting.
\newblock In {\em Proceedings of the Platform for Advanced Scientific Computing
  Conference}, PASC '18, New York, NY, USA, 2018. Association for Computing
  Machinery.

\bibitem{Kwak10www}
Haewoon Kwak, Changhyun Lee, Hosung Park, and Sue Moon.
\newblock {W}hat is {T}witter, a social network or a news media?
\newblock In {\em WWW '10: Proceedings of the 19th international conference on
  World wide web}, pages 591--600, New York, NY, USA, 2010. ACM.

\bibitem{maley2019conveyors}
F.~Maley and Jason DeVinney.
\newblock Conveyors for streaming many-to-many communication.
\newblock In {\em 2019 IEEE/ACM 9th Workshop on Irregular Applications:
  Architectures and Algorithms (IA3)}, pages 1--8, 11 2019.

\bibitem{meusel2014webcc}
Robert Meusel, Sebastiano Vigna, Oliver Lehmberg, and Christian Bizer.
\newblock Graph structure in the web---revisited: a trick of the heavy tail.
\newblock In {\em WWW Companion}, pages 427--432, 2014.

\bibitem{neo4j.property.graph}
Neo4j.
\newblock Neo4j - property graph.
\newblock https://neo4j.com/developer/graph-database/\#property-graph, 2016.

\bibitem{Orientdb.001}
OrientDB.
\newblock Orientdb.
\newblock https://orientdb.com, 2018.

\bibitem{Pandey.2019.HPEC.H-INDEX.8916492}
S.~{Pandey}, X.~S. {Li}, A.~{Buluc}, J.~{Xu}, and H.~{Liu}.
\newblock H-index: Hash-indexing for parallel triangle counting on {GPU}s.
\newblock In {\em 2019 IEEE High Performance Extreme Computing Conference
  (HPEC)}, pages 1--7, 2019.

\bibitem{Pandey.2021.TRUST.9373989}
S.~Pandey, Z.~Wang, S.~Zhong, C.~Tian, B.~Zheng, X.~Li, L.~Li, A.~Hoisie,
  C.~Ding, D.~Li, and H.~Liu.
\newblock Trust: Triangle counting reloaded on {GPU}s.
\newblock {\em IEEE Transactions on Parallel \& Distributed Systems},
  (01):1--1, mar 5555.

\bibitem{benson2017temporal}
Ashwin Paranjape, Austin~R. Benson, and Jure Leskovec.
\newblock Motifs in temporal networks.
\newblock In {\em Proceedings of the Tenth ACM International Conference on Web
  Search and Data Mining}, WSDM '17, pages 601--610, New York, NY, USA, 2017.
  Association for Computing Machinery.

\bibitem{Pearce.2019.HPEC.DistributedTriangle.8916243}
R.~{Pearce}, T.~{Steil}, B.~W. {Priest}, and G.~{Sanders}.
\newblock One quadrillion triangles queried on one million processors.
\newblock In {\em 2019 IEEE High Performance Extreme Computing Conference
  (HPEC)}, pages 1--5, 2019.

\bibitem{pearce2017triangle}
Roger Pearce.
\newblock Triangle counting for scale-free graphs at scale in distributed
  memory.
\newblock In {\em 2017 IEEE High Performance Extreme Computing Conference
  (HPEC)}, pages 1--4. IEEE, 2017.

\bibitem{Pearce:2014:FPT:2683593.2683654}
Roger Pearce, Maya Gokhale, and Nancy~M. Amato.
\newblock Faster parallel traversal of scale free graphs at extreme scale with
  vertex delegates.
\newblock In {\em Proceedings of the International Conference for High
  Performance Computing, Networking, Storage and Analysis}, SC '14, pages
  549--559, Piscataway, NJ, USA, 2014. IEEE Press.

\bibitem{Priest2019Grapl}
B.~{Priest}, T.~{Steil}, R.~{Pearce}, and G.~{Sanders}.
\newblock You've got mail (ygm): Building missing asynchronous communication
  primitives.
\newblock In {\em 2019 IEEE International Parallel and Distributed Processing
  Symposium Workshops (IPDPSW)}, page~2, May 2019.

\bibitem{reza2020indexing}
Tahsin Reza, Matei Ripeanu, Geoffrey Sanders, and Roger Pearce.
\newblock Labeled triangle indexing for efficiency gains in distributed
  interactive subgraph search.
\newblock In {\em 2020 IEEE/ACM 10th Workshop on Irregular Applications:
  Architectures and Algorithms (IA3)}. IEEE Computer Society, nov 2020.

\bibitem{Reza:2018:PPT:3291656.3291684}
Tahsin Reza, Matei Ripeanu, Nicolas Tripoul, Geoffrey Sanders, and Roger
  Pearce.
\newblock Prunejuice: Pruning trillion-edge graphs to a precise
  pattern-matching solution.
\newblock In {\em Proceedings of the International Conference for High
  Performance Computing, Networking, Storage, and Analysis}, SC '18, pages
  21:1--21:17, Piscataway, NJ, USA, 2018. IEEE Press.

\bibitem{Rossi2015network_repository}
Ryan~A. Rossi and Nesreen~K. Ahmed.
\newblock The network data repository with interactive graph analytics and
  visualization.
\newblock In {\em AAAI}, 2015.

\bibitem{Roy:2015:CSG:2815400.2815408}
Amitabha Roy, Laurent Bindschaedler, Jasmina Malicevic, and Willy Zwaenepoel.
\newblock Chaos: Scale-out graph processing from secondary storage.
\newblock In {\em Proceedings of the 25th Symposium on Operating Systems
  Principles}, SOSP '15, pages 410--424, New York, NY, USA, 2015. ACM.

\bibitem{kepner2017summary}
Siddharth Samsi, Vijay Gadepally, Michael~B. Hurley, Michael Jones, Edward~K.
  Kao, Sanjeev Mohindra, Paul Monticciolo, Albert Reuther, Steven~Thomas Smith,
  William Song, Diane Staheli, and Jeremy Kepner.
\newblock Graph{C}hallenge.org: Raising the bar on graph analytic performance.
\newblock {\em CoRR}, 2018.

\bibitem{kepner2019summary}
Siddharth Samsi, Jeremy Kepner, Vijay Gadepally, Michael Hurley, Michael Jones,
  Edward Kao, Sanjeev Mohindra, Albert Reuther, Steven Smith, William Song,
  Diane Staheli, and Paul Monticciolo.
\newblock Graph{C}hallenge.org triangle counting performance, 2020.

\bibitem{Schank2007_1000007159}
Thomas Schank.
\newblock {\em Algorithmic Aspects of Triangle-Based Network Analysis}.
\newblock PhD thesis, 2007.

\bibitem{Serafini:2017:QDG:3127479.3131625}
Marco Serafini, Gianmarco De~Francisci~Morales, and Georgos Siganos.
\newblock Qfrag: Distributed graph search via subgraph isomorphism.
\newblock In {\em Proceedings of the 2017 Symposium on Cloud Computing}, SoCC
  '17, pages 214--228, New York, NY, USA, 2017. ACM.

\bibitem{friendster}
Friendster social network.
\newblock Friendster: The online gaming social network.
\newblock {https://archive.org/details/friendster-dataset-201107}.

\bibitem{Sundaram:2015:GHP:2809974.2809983}
Narayanan Sundaram, Nadathur Satish, Md~Mostofa~Ali Patwary, Subramanya~R.
  Dulloor, Michael~J. Anderson, Satya~Gautam Vadlamudi, Dipankar Das, and
  Pradeep Dubey.
\newblock Graphmat: High performance graph analytics made productive.
\newblock {\em Proc. VLDB Endow.}, 8(11):1214--1225, July 2015.

\bibitem{Suri:2011:CTC:1963405.1963491}
Siddharth Suri and Sergei Vassilvitskii.
\newblock Counting triangles and the curse of the last reducer.
\newblock In {\em Proceedings of the 20th International Conference on World
  Wide Web}, WWW '11, pages 607--614, New York, NY, USA, 2011. ACM.

\bibitem{Teixeira:2015:ASD:2815400.2815410}
Carlos H.~C. Teixeira, Alexandre~J. Fonseca, Marco Serafini, Georgos Siganos,
  Mohammed~J. Zaki, and Ashraf Aboulnaga.
\newblock Arabesque: A system for distributed graph mining.
\newblock In {\em Proceedings of the 25th Symposium on Operating Systems
  Principles}, SOSP '15, pages 425--440, New York, NY, USA, 2015. ACM.

\bibitem{TigerGraph.001}
TigerGraph.
\newblock Tigergraph.
\newblock https://www.tigergraph.com, 2018.

\bibitem{Tom.2019.ICPP.DistributedTriangle.10.1145/3337821.3337853}
Ancy~Sarah Tom and George Karypis.
\newblock A 2d parallel triangle counting algorithm for distributed-memory
  architectures.
\newblock In {\em Proceedings of the 48th International Conference on Parallel
  Processing}, ICPP 2019, New York, NY, USA, 2019. Association for Computing
  Machinery.

\bibitem{Wang.2019.HPEC.GPUTriangle.8916434}
L.~{Wang} and J.~D. {Owens}.
\newblock Fast bfs-based triangle counting on gpus.
\newblock In {\em 2019 IEEE High Performance Extreme Computing Conference
  (HPEC)}, pages 1--6, 2019.

\bibitem{wolf2015}
M.~M. Wolf, J.~W. Berry, and D.~T. Stark.
\newblock A task-based linear algebra building blocks approach for scalable
  graph analytics.
\newblock In {\em 2015 IEEE High Performance Extreme Computing Conference
  (HPEC)}, pages 1--6, Sept 2015.

\bibitem{Yasar.2019.HPEC.Triangle.8916233}
A.~{Ya\c{s}ar}, S.~{Rajamanickam}, J.~{Berry}, M.~{Wolf}, J.~S. {Young}, and
  {\"{U}}.~V. {\c{C}ataly\"{u}rek}.
\newblock Linear algebra-based triangle counting via fine-grained tasking on
  heterogeneous environments : (update on static graph challenge).
\newblock In {\em 2019 IEEE High Performance Extreme Computing Conference
  (HPEC)}, pages 1--4, 2019.

\bibitem{yasar2020blockbased}
Abdurrahman Yaşar, Sivasankaran Rajamanickam, Jonathan Berry, and Ümit
  V.~Çatalyürek.
\newblock A block-based triangle counting algorithm on heterogeneous
  environments, 2020.

\bibitem{Zhang.2018.HPEC.SharedTriangle.8547569}
J.~{Zhang}, D.~G. {Spampinato}, S.~{McMillan}, and F.~{Franchetti}.
\newblock Preliminary exploration of large-scale triangle counting on
  shared-memory multicore system.
\newblock In {\em 2018 IEEE High Performance extreme Computing Conference
  (HPEC)}, pages 1--6, 2018.

\bibitem{zhang2019litete}
Yongxuan Zhang, Hong Jiang, Fang Wang, Yu~Hua, Dan Feng, and Xianghao Xu.
\newblock Litete: Lightweight, communication-efficient distributed-memory
  triangle enumerating.
\newblock {\em IEEE Access}, 7:26294--26306, 2019.

\end{thebibliography}
\bibliographystyle{plain}

\end{document}